\documentclass[usenatbib]{mnras}
\usepackage{amsmath,amssymb}
\usepackage{graphicx}

\def\vb#1{\mbox{\boldmath $#1$}}

\newcommand{\grackle}{{\sc grackle}}
\newcommand{\enzo}{{\sc enzo}}

\newcommand{\percc}{{\rm cm^{-3}}}
\newcommand{\um}{{\rm \mu m}}
\newcommand{\E}[1]{\times 10^{#1}}

\newcommand{\nH}{n_{{\rm H}}}
\newcommand{\mH}{m_{{\rm H}}}

\newcommand{\Mhalo}{M_{\rm halo}}
\newcommand{\Rhalo}{R_{\rm halo}}
\newcommand{\zform}{z_{\rm form}}

\newcommand{\MPopIII}{M_{\rm Pop III}}
\newcommand{\Esn}{E_{{\rm SN}}}

\newcommand{\tff}{t_{\rm ff}}
\newcommand{\tcol}{t_{\rm col}}
\newcommand{\tlife}{t_{\rm life}}

\newcommand{\QH}{Q({\rm H})}
\newcommand{\QHH}{Q({\rm H}_2)}

\newcommand{\cs}{c_{\rm s}}

\newcommand{\abFe}[1]{{\rm [{#1}/Fe]}}

\newcommand{\Enstatite}{{\rm MgSiO_3}}
\newcommand{\Forsterite}{{\rm Mg_2SiO_4}}

\newcommand{\Silica}{{\rm SiO_2}}
\newcommand{\Troilite}{{\rm FeS}}

\newcommand{\Magnesia}{{\rm MgO}}

\newcommand{\namb}{n_{{\rm amb}}}

\newcommand{\XH}{X_{{\rm H}}}
\newcommand{\Zsun}{{\rm Z_{\bigodot}}}
\newcommand{\Msun}{{\rm M_{\bigodot}}}

\defcitealias{Chiaki15}{C15}
\defcitealias{Chiaki16}{C16}
\defcitealias{Chiaki18}{C18}
\defcitealias{Smith15}{S15}
\defcitealias{Hirano14}{H14}
\defcitealias{Ishigaki14}{I14}
\defcitealias{Tominaga14}{T14}
\defcitealias{Marassi15}{M15}

\usepackage{color}
\definecolor{rev}{rgb}{0.8,0.0,0.0}

\title[Seeding the Second Star]
      {Seeding the second star: Enrichment from population III, dust evolution, and cloud collapse}

\author[G. Chiaki and J.~H. Wise]
{Gen Chiaki$^{1,2}$\thanks{E-mail: gen.chiaki@physics.gatech.edu} and
John H. Wise$^{1}$
\\
$^{1}$Center for Relativistic Astrophysics, School of Physics, Georgia Institute of Technology, Atlanta, GA 30332, USA \\
$^{2}$Department of Physics, Konan University,
8-9-1 Okamoto, Kobe, 658-0072, Japan}

\begin{document}

\date{}

\pagerange{\pageref{firstpage}--\pageref{lastpage}} \pubyear{2018}

\maketitle

\label{firstpage}

\begin{abstract}
We investigate the formation of extremely metal-poor (EMP) stars that are observed in the Galactic 
halo and neighboring ultra-faint dwarf galaxies.
Their low metal abundances (${\rm [Fe/H]} < -3$) indicate that their parent clouds were enriched
by a single or several supernovae (SNe) from the first (Pop III) stars.
In this study, we perform numerical simulations of the entire formation sequence of a
EMP star through the feedback effects of photo-ionization and metal-enrichment
by a Pop III SN.
We for the first time employ a metal/dust properties calculated consistently with the progenitor model,
and solve all relevant radiative cooling processes and chemical reactions including
metal molecular formation and grain growth until the protostar formation.
In a minihalo (MH) with mass $1.77\E{6} \ \Msun$, a Pop III star with mass $13 \ \Msun$
forms at redshift $z=12.1$.
After its SN explosion, the shocked gas falls back into the central MH internally enriching itself.
The metallicity in the recollapsing region is $2.6\E{-4} \ \Zsun$ (${\rm [Fe/H]} = -3.42$).
The recollapsing cloud undergoes cooling by HD, CO, and OH molecules and heating along with H$_2$ formation.
Eventually by grain growth and dust cooling, knotty filaments appear in the central 100 au
region with the help of turbulence driven by the SN, leading to the formation of low-mass EMP stars
surviving until the present day.
\end{abstract}



\section{INTRODUCTION}

The properties of early-generation stars are 
crucial to determine structure formation in the first billion
years from Big Bang including the first galaxy formation
and cosmic reionization.
The typical mass scale and initial mass function (IMF) of
first-generation metal-free stars (Pop III stars) determine
the degrees of photoionization and the metal enrichment in
the ambient interstellar medium (ISM) through
their supernova (SN) explosions.
Conversely, the metal content in collapsing clouds 
affects the mass range of stars formed \citep{Omukai00, Bromm03}.
Such co-evolution determines the star formation history of the
ISM in high-redshift galaxies 
\citep[see a recent review of][and references therein]{Dayal18Rev}.

In the next decade, instruments such as {\it James-Webb Space Telescope} ({\it JWST}),
{\it SPICA}, and {\it Thirty-Meter Telescope} ({\it TMT}) are planned to directly observe 
high-redshift galaxies to extend our knowledge of their IMFs and metal content.
The prediction of the observational features of high-redshift galaxies and even the first generation
of galaxies is an urgent task
\citep[see a review of observational features of {\it JWST} in][]{Kalirai18}.
On the theoretical side, cosmological simulations of the first stars and galaxies,
including massive star feedback, provide valuable insight of their origin and evolution
\citep{Tumlinson07, Salvadori07, Wise12, Barrow17, Barrow18, Magg17, Dayal18BH}.

In this work, we focus on another constraint on IMFs
in the early Universe.
In the halo and bulge regions in the Milky Way galaxy and the neighboring ultra-faint
dwarf (UDF) galaxies, ancient stars with metallicities lower than the solar value,
so-called metal-poor stars, have been observed.
In particular, ones with iron abundances of $\rm [Fe/H] < -3$ are called extremely
metal-poor (EMP) stars \citep{Beers05}.
Due to their small metal content, they should inherit the nucleosynthetic features
of a single or several parent Pop III SN(e) \citep{Ryan96, Cayrel04, Chiaki18}.
Constraining the mass scale and explosion mechanism of Pop III stars
indirectly through the elemental abundance ratios and metallicities of metal-poor stars
is called near-field cosmology or Galactic archaeology \citep{Ishigaki18}.

The second generation of star formation is the very first matter cycle in the ISM.
It begins with Pop III star formation in small gravitationally-bound dark matter
(DM) haloes, so-called minihaloes (MHs), with masses $\sim 10^6 \ \Msun$ at redshift $z\sim 20$.
The first light is emitted during its main sequence, and the first metals are then
synthesized and released by its SN if it is massive \citep[$\MPopIII = 8$--$40 \ \Msun$
and $140$--$260 \ \Msun$;][]{Heger02}.
Metals are incorporated into later star-forming regions, where
the cloud collapses and fragments into low-mass clumps ($\sim 0.01$--$0.1 \ \Msun$)
by gas cooling of dust thermal emission 
into the embryos of observed long-lived EMP stars \citep{Omukai00, Schneider03, Bromm14}.

There are two processes of metal enrichment.
One is called external enrichment (EE), where the gas shocked by a Pop III SN reaches a 
neighboring clump and triggers star formation.
The other is called internal enrichment (IE), in which
the metal-enriched gas falls back into the DM potential having hosted
the Pop III star.
\citet[][\citetalias{Chiaki18}]{Chiaki18} showed that both processes occur through a balance between the
radiation/supernova energy by a Pop III star and binding energy of MH,
and that the latter is the major process for the formation of EMP stars.

The whole process has only been investigated by \citet{Smith15}.
They show that an externally enriched clump with metallicity $\sim 2\E{-5} \ \Zsun$ contracts
to fragment by dust cooling.
Although IE is the major process for the formation of EMP stars \citepalias{Chiaki18},
its whole process has not been explicitly followed by three-dimensional numerical simulations
in the cosmological context.
\citet{Machida05} and \citet{Chiaki13} studied the low-mass star formation in the partially enriched
SN shell with one-zone and one-dimensional calculations.
However, because the propagation of SN shells is highly anisotropic, three-dimensional simulations
are required to properly follow the process of metal mixing and gas fragmentation 
\citep{Ritter12}.
The works of \citet{Ritter12, Ritter15, Ritter16}, \citet{Sluder16}, and \citetalias{Chiaki18} are
limited in the sense that they only consider the recollapse of the shocked gas, i.e., the succeeding 
protostar formation and cloud fragmentation are not followed.

In low-metallicity clouds, various heating/cooling processes are present.
\citet[][\citetalias{Chiaki16}]{Chiaki16} followed the cloud collapse with 
metallicities $10^{-6}$--$10^{-3} \ \Zsun$.
They showed that rapid gas heating along with H$_2$ molecular formation
prevents fragmentation in the most cases even though the dust cooling is efficient later.
In other cases where OH/H$_2$O molecular cooling operates, 
fragmentation along the filamentary structure appears.
In \citetalias{Chiaki16}, metals are uniformly added in static clouds.
In the shock-driven cloud collapse, the dynamical timescale becomes shorter
so that the efficiencies of these heating/cooling processes might be suppressed.
Furthermore, the turbulence driven by SN shocks will enhance the fragmentation \citep{Dopcke11, Dopcke13}.
To properly know how the hydrodynamic evolution is coupled with the chemical processes,
it is required to follow the cloud collapse with full calculations of chemistry and gas
heating/cooling in these shock-driven star-forming clouds.

The metal and dust properties produced from a Pop III SN are a key ingredient in these calculations
\citep{Bovino16, Grassi17}.
In the recollapsing region, the metal and dust compositions and dust size distribution
are assumed to directly inherit the nucleosynthesis and nucleation in Pop III SN ejecta \citep{Tanaka17}.
Type-II SNe yield more abundant alpha elements $\abFe{\alpha} > 0$ ($\alpha = {\rm O},~{\rm Mg},~{\rm Si}$)
than measured in the present-day including the contribution of Type-Ia SNe \citep{Umeda02}.
This elemental abundance enhances O {\sc i} fine-structure cooling and OH/H$_2$O molecular cooling.
Although dust grains are formed in expanding ejecta, they are partly destroyed by reverse shock,
and return to the gas phase \citep{Bianchi07, Nozawa07}.
Consequently, the condensation efficiency of metal (dust to metal mass ratio) is smaller (a few \%)
than in the present-day (50\%), where all refractory elements (Mg, Si, and Fe) are locked
up into grains \citep{Pollack94}.
On the other hand, grains can grow by accreting the gas-phase metal (grain growth) in a molecular
clouds, which enhances the dust cooling efficiency \citep[][\citetalias{Chiaki15}]{Chiaki15}.

In this work, we for the first time to follow the full process from Pop III to Pop II
star formation via IE in a three-dimensional cosmological simulation
with all relevant chemistry and heating/cooling processes.
We calculate the metal and dust models employed in this study with accurate nucleosynthesis and nucleation models
of a Pop III SN.
We describe our numerical models in Sec. \ref{sec:method} and the results in Sec. \ref{sec:results}.
Then, we discuss importance of physical processes included in this work and caveates 
(Sec. \ref{sec:discussion}).
The application of our results to Galactic Archaeology is discussed in Sec. \ref{sec:GA}.
Finally, we provide summary and conclusion of this work in Sec. \ref{sec:conclusion}.


\section{Numerical models}
\label{sec:method}

\subsection{Simulation setup}

We use the $N$-body/adaptive mesh refinement (AMR) cosmological
hydrodynamics simulation code \enzo~\citep{Bryan14}.\footnote{\url{http://enzo-project.org/}}
To follow the evolution of shocks with high Mach numbers, a grid-based code is useful relative
to smoothed particle hydrodynamics (SPH) codes by which the spurious surface tension affects 
the hydrodynamics \citep[also see][]{Saitoh13, Hopkins13, Rosswog15}.
The code solves the hydrodynamics equations with the Piecewise
Parabolic Method (PPM) in an Eulerian frame while using a two-shock Riemann solver.
In the hydrodynamics equations of {\sc enzo} PPM solver, the fixed adiabatic index $\gamma = 5/3$
has been imposed.
At densities $\nH \gtrsim 10^8 \ \percc$, the gas becomes fully molecular at low metallicities
$Z<10^{-3} \ \Zsun$, and
$\gamma$ approaches 7/5.
If $\gamma$ remains 5/3 in the molecular regime, this causes spurious heating.
To avoid this, we have modified the hydrodynamics solver to consider a variable $\gamma$ value.

For the study of star formation and expansion of dense
cooling shells, the AMR technique can capture the inherently wide
dynamic range of length and density of this problem.  We adopt the
following cosmological parameters in the simulation: $\Omega _{\rm
  m} = 0.3089$, $\Omega _{\rm CDM} = 0.2603$, $\Omega _{\rm \Lambda}
= 0.6911$, and $H_0 = 67.74 \ {\rm km \ s^{-1} \ Mpc^{-1}}$
\citep{Planck2015}.  We initialize the periodic simulation volume
with a side length of 300 comoving kpc at redshift $z=140$ with
\textsc{MUSIC} \citep{Hahn11}.  We initially perform a small $64^3$
pathfinder simulation with only dark matter to identify the most
massive halo at $z=10$ that has a mass of $3.8 \times 10^6~\Msun$.
We then reinitialize and recenter the simulation by resampling its
Lagrangian volume at $z=10$ increasing the mass and spatial
resolution by a factor of 64 and 4, respectively, i.e. two
additional initial AMR levels.  The Lagrangian volume considered is
a sphere centered on the target halo with a radius of five times the
virial radius.  In the ``zoom-in'' region, we have an effective
initial resolution of $256^3$ and a mass resolution of $53.4~\Msun$,
resolving the halo by approximately 50,000 particles.  Although this
simulation size is smaller than most cosmological simulations of
metal-free and metal-poor star formation, it was necessary to reduce
its computational size because of the expense of the solving a
detailed chemical network that includes a live dust model.  In the
end, the full simulation required 33 days on 224 cores, consuming
177,000 core-hours.  We terminate the simulations when the first
metal-enriched hydrostatic core forms at a density of
$10^{16}~\percc$.  We perform all of the analysis in this paper with
the {\tt yt} toolkit \citep{yt}.

Computational cells satisfying criteria below are progressively
refined by a factor of two in space up to 33 refinement levels,
resulting in a maximal spatial resolution of 0.01~au.  We smooth the
dark matter density field at scales below 1 comoving pc (AMR level
12).  At these scales, the gas density dominates, and by smoothing the
dark matter density, we remove any artifacts, i.e. heating in the wake
of individual massive particles, associated with the discrete
representation of the dark matter mass distribution
\citep[e.g.][]{Abel02}.  We employ the following refinement criteria:
\begin{itemize}
\item[(i)] the baryon mass in a cell on a refinement level $l$ exceeds
$3\times 2 ^{-0.2 l}$ times the mean baryon mass on the root grid, 
\item[(ii)] the DM particle mass in a cell exceeds
three times the initial mass, 
\item[(iii)] the local Jeans length is less than 64 times the cell size. 
\end{itemize}

We allow Pop III star formation in regions where a cell satisfying the
following criteria \citep{Wise12}:
\begin{itemize}
\item[(i)] the physical gas density exceeds $10^6 \ \percc$,
\item[(ii)] the gas flow is convergent ($\nabla \cdot \vb{v} < 0$),
\item[(iii)] the cooling time is less than the dynamical time,
\item[(iv)] the metallicity is less than the critical value ($5\E{-5} \ \Zsun$),
\item[(v)] the H$_2$ fraction exceeds a critical value ($10^{-3}$) that is typical for the collapsing primordial cloud at the critical density (i).
\end{itemize}
Because we focus on the feedback effects of a single Pop III star in this work,
the formation of secondary Pop III stars is artificially prohibited.
The mass of a Pop III star is sampled from the IMF of
\begin{equation}
f(\log \MPopIII) = M^{-1} \exp \left[ -\left( \frac{M_{\rm char}}{\MPopIII } \right)^{1.6} \right],
\label{eq:PopIII_IMF}
\end{equation}
where $M_{\rm char}$ is the characteristic mass of Pop III stars, and $M_{\rm char} = 20 \ \Msun$.
We regard the star particle as a radiation point source, and solve the radiative transfer equation with \textsc{Moray} \citep{Wise11} during the lifetime
$\tlife$ of the star.
Then the supernova energy of $\Esn = 10^{51}$ erg and metal is added within the radius of 10 pc.
Although the stellar core is stratified just before the explosion, where heavier elements such as
iron reside inner region, we assume uniform i.e. fully-mixed ejecta, considering hydrodynamical instabilities 
\citep{Chen17b}.
Metals are initially contained in the ejecta surrounded by the ambient medium
through a contact discontinuity (CD) that is placed at a radius of 7.5
pc, which is appropriate for the energy-conserving phase.
The metals can penetrate through this barrier into the ambient gas
when the shock undergoes Rayleigh-Taylor (RT) instabilities on the CD.

\subsection{Chemistry and cooling}

We use the chemistry/cooling library \grackle, generalized
to adopt simulation codes with various schemes 
\citep{Smith17}.\footnote{\url{https://grackle.readthedocs.io/}}
The original version presented by \citet{Smith17} solved the chemical 
reactions of at most 12 primordial species and radiative cooling of the relevant 
species.
The metal cooling rates were interpolated from tables calculated by {\sc cloudy} \citep{Smith08}.
The dust temperature and cooling efficiency were solved based on the dust
model in the interstellar medium (ISM) in the vicinity of the solar system \citep{Pollack94, Omukai00}.
\grackle~is well-designed to supplement the default chemical reactions and cooling functions with user-defined ones.
In order to follow the abundances and cooling efficiencies of 48 chemical species of metal elements
and dust as well as primordial elements 
for a wide range of density ($\nH < 10^{18} \ \percc$) and gas temperature ($T < 10^9$ K),
we add the corresponding chemical reactions, cooling functions, and opacities described in the
subsequent sections.
We also modified the scheme to solve the ordinary differential equations of reaction rates 
from the backward differential formula \citep{Anninos97} to an implicit scheme.

\subsubsection{Primordial chemistry and cooling}
We extend {\sc grackle}'s chemical network of primordial species to include 49 reactions
for 15 primordial species: 
$\rm e$, $\rm  H$, $\rm  H^+$, $\rm  H_2$, $\rm  H^-$, $\rm H_2^+$,
$\rm HeH^+$, $\rm  He$, $\rm  He^+$, $\rm  He^{2+}$,
$\rm D$, $\rm  D^+$, $\rm  D^-$, $\rm  HD$, and $\rm  HD^+$ \citepalias{Chiaki18}
by adding $\rm HeH^+$, $\rm D^-$, and $\rm HD^+$, which are important in the high-redshift intergalactic
medium \citep[$z>10$;][]{Galli13}.
The reactions include the collisional ionization/recombination and formation/destruction of H$_2$ via
H$^-$/H$_2^+$ processes and three-body reactions.

We include gas heating along with H$_2$ molecular formation, which is important to suppress
the gas elongation and fragmentation \citepalias{Chiaki16}.
For $T>8000$ K, we consider bremsstrahlung, inverse Compton cooling, ionization/recombination and
transition line cooling of H- and He-bearing species.  
For $T<10000$ K, we calculate the molecular cooling rates by H$_2$ and HD.
The radiative cooling rate of H$_2$ molecules are calculated for each line transition
between 20 rotational and 3 vibrational levels.
For HD, we consider 3 vibrational levels.

For molecular cooling, the emission rate is reduced by the corresponding photon escape fraction 
$\beta _{\rm line} = \frac{1-e^{-\tau _{\rm line}}}{\tau _{\rm line}} e^{-\tau _{\rm cont}}$,
where $\tau _{\rm line} = \alpha _{\rm line} l_{\rm sh}^{\rm line}$ is the optical depth for
each transition line.
In this formula, $\alpha _{\rm line}$ is the absorption coefficient including stimulated
emission, and
$l_{\rm sh} = \min \{ 2v_{\rm th} / (dv/dr),~l_{\rm Jeans} \}$ is the shielding length as the minimum 
between the Sobolev length, where the relative velocity of the fluid becomes equivalent to
line width of the order of thermal velocity $v_{\rm th} = (2kT/m_j)^{1/2}$,
and the Jeans length \citep{Takahashi83}.
For simplicity, we here estimate the velocity gradient as $dv/dr = 1/3t_{\rm ff}$ under the one-zone
approximation \citep{Omukai00}.
We will present the formulation of continuum opacity $\tau _{\rm cont}$ in Section \ref{sec:dust_physics}.

Throughout the simulation, hydrogen mass fraction is a constant $\XH = 0.76$.
Although the helium fraction is enhanced just around a SN ejecta, the He-rich gas
is diluted by the ambient pristine gas.
The density is expressed as the number density of hydrogen nuclei
$\nH = \XH \rho /\mH$ with mass density $\rho$ throughout this paper.

\subsubsection{Metal chemistry and cooling}
The metal cooling is important to determine the fragmentation properties of low-metallicity gas
in the early stage of collapse \citep{Bromm03}.
In particular, metal molecules such as OH and H$_2$O
can overcome heating associated with H$_2$ formation to enhance 
fragmentation \citepalias{Chiaki16}.\footnote{Although CO is important in the present-day
ISM, not in the low-metallicity ISM affected only by massive Pop III SNe \citepalias[$\MPopIII > 20 \ \Msun$;][]{Chiaki16}.}
We for the first time follow the abundances of their different phases (ions, atoms, and molecules)
in three-dimensional simulations of Pop III SNRs,
considering 40 reactions for the 19 species:
C$^+$, C, CH, CH$_2$, CO$^+$, CO, CO$_2$, O$^+$, O, OH$^+$, OH,
H$_2$O$^+$, H$_2$O, H$_3$O$^+$, O$_2^+$, O$_2$, Si, SiO, and SiO$_2$.
We solve recombination of C$^+$ and formation of molecules CH, CH$_2$, CO, and CO$_2$.
For O-bearing species, recombination/charge transfer of O$^+$ and OH, O$_2$, and H$_2$O formation
is included.
Atoms of Si are oxidized into SiO and SiO$_2$ molecules in the course of collapse \citepalias{Chiaki16}.

The fine-structure level transition line cooling of C$^+$, C and O for the systems of
2, 3, and 5 levels are considered, respectively \citep{Santoro06}.
The photon escape fractions are estimated in the same manner as H$_2$ and HD.
We calculate the cooling rates by rotational level transition of OH, H$_2$O, and CO by interpolating the
tables of \citet{Neufeld93}, \citet{Neufeld95}, and \citet{Omukai10}, respectively.

\subsubsection{Dust physics}
\label{sec:dust_physics}

Dust grains play a crucial role in determining the final fate of
a cloud by enhancing gas cooling efficiency and thus inducing
fragmentation \citep[e.g.][\citetalias{Chiaki16}]{Smith15, SafranekShrader16}.
They modify the thermodynamics of gas in the four aspects:
\begin{enumerate}
\item[] (1) H$_2$ molecular formation on grain surfaces,
\item[] (2) continuum opacity,
\item[] (3) gas cooling by dust thermal emission, and
\item[] (4) accretion of gas-phase metal onto grains \citepalias[grain growth:][]{Chiaki15}.
\end{enumerate}
To self-consistently consider the properties of dust grains released by
Pop III SNe, we derive these efficiencies separately with grain radii
$r = 1 \ $\AA--$10 \ \um$ with 0.1 dex interval
for eight grain species:
metallic silicon (Si),
metallic iron (Fe),
forsterite ($\Forsterite$),
enstatite  ($\Enstatite $),
amorphous carbon (C),
silica     ($\Silica    $),
magnesia   ($\Magnesia  $), and
troilite   ($\Troilite  $).

The continuum opacity is estimated as 
$\tau _{\rm cont} = (\kappa _{\rm p} \rho + \sum _i \kappa _{\rm i} \rho _i) l_{\rm Jeans}$,
where $\kappa _{\rm p}$ and $\kappa _i$ are the Planck mean opacity of the primordial gas \citep{Mayer05}
and a grain species $i$ \citep{Nozawa08}, and 
$\rho _i$ is the mass density of a grain species $i$.
We use the Jeans length as the shielding length for simplicity.
The continuum cooling rates such as collision-induced emission (CIE) and dust thermal emission
are reduced by a factor of $\beta _{\rm cont} = \min \{1, \tau _{\rm cont} ^{-2}\}$
under the diffusion approximation.

For densities $\nH \gtrsim 10^{10} (Z/10^{-4} \ \Zsun)^{-2} \ \percc$,
grain growth is present and can enhance cloud fragmentation \citepalias{Chiaki16}.
We simultaneously compute the loss of elements Mg, Al, S, and Fe, assuming
that they all are in the form of atoms.
Carbon grains grow through impact of carbon atoms.
Since C atoms are rapidly oxidized to form CO molecules at density $\nH \sim 10^6 \ \percc$,
growth of carbon grains can occur if the number abundance ratio $\rm C/O > 1$.
Silicate grains grow by accreting Mg, SiO, and H$_2$O molecules.
Since the dominant Si-bearing species at $\nH > 10^6 \ \percc$ is SiO, 
we expect that silicate grains grow in the Pop III SN dust models.

The formulation of the rates of (1)--(4) explicitly described
in our previous paper \citepalias{Chiaki15} depends on the grain temperature.
To calculate it, many iterations are required to solve the energy balance equation
between gas-dust heat transfer rate and dust thermal emission rate for each grain
species and size bin.
In order to reduce the computational costs, 
these rates have been tabulated for each grain species,
and are interpolated from  
density ($\rho$), gas temperature ($T$), density of a grain species $i$ ($\rho _i$),
and density of the corresponding key element $X$ ($\rho_{X}$) of a fluid cell at each timestep.
In this strategy, we successfully include the effect of grain growth 
(See Appendix \ref{sec:rates_dust} for the detailed description).

\begin{figure}
\includegraphics[width=\columnwidth]{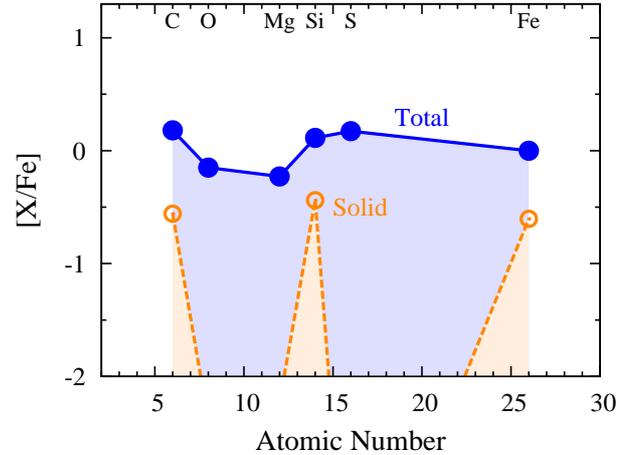}
\caption{
{\it Blue solid line}: abundance $[X/{\rm Fe}]$ of major elements $X$ relative to
iron for our SN model with progenitor mass $\MPopIII = 13 \ \Msun$.
{\it Orange dashed line}: number fraction of nuclei condensed into dust grains
for our dust formation/destruction model with ambient gas density $\namb = 1 \ \percc$. 
For metallicity $Z_{\rm recol} = 2.6\E{-4} \ \Zsun$ in the recollapsing region (see text),
the absolute iron abundance is ${\rm [Fe/H]} = -3.42$.
}
\label{fig:metal}
\end{figure}

\begin{figure}
\includegraphics[width=\columnwidth]{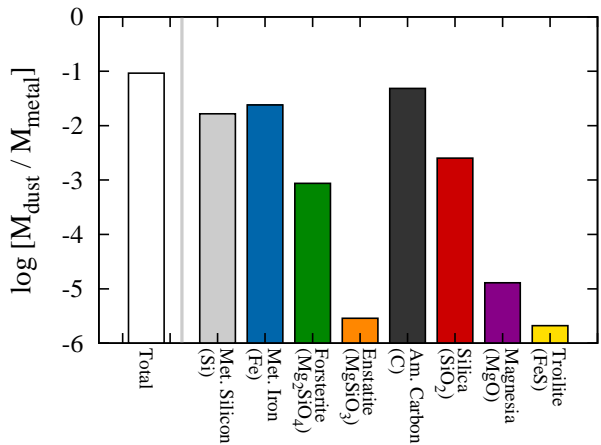}
\caption{
Mass fraction of the total and individual dust grain species relative to all metals after the dust destruction
by reverse shocks in the supernova remnant.
}
\label{fig:dust}
\end{figure}

\begin{figure}
\includegraphics[width=\columnwidth]{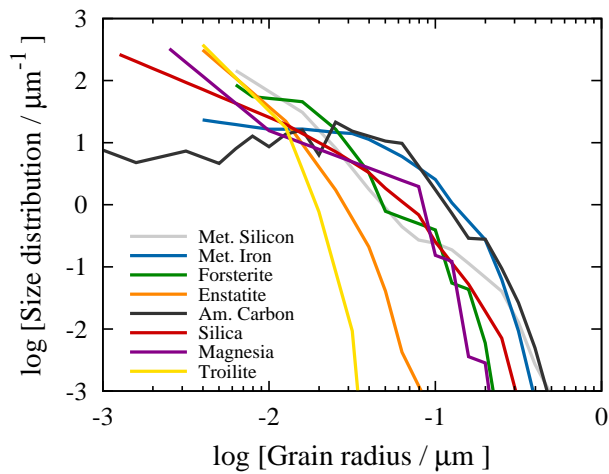}
\caption{
Size distribution function of each dust grain species normalized to unity just after they are
ejected from the Pop III supernova.
}
\label{fig:dist}
\end{figure}

\subsection{Pop III supernova yields}

In this paper, we employ the Pop III SN model
in a hydrodynamic evolution of a SN remnant and
star-forming clouds self-consistently.
In this simulation, a Pop III star with mass $13.5 \ \Msun$ forms.
Metal and dust abundances and dust size distribution
are taken from \citet{Umeda02} and \citet{Nozawa03, Nozawa07} 
for the progenitor mass $\MPopIII = 13 \ \Msun$, 
explosion energy $\Esn = 1\E{51}$ erg, and
ambient gas density $\namb = 1 \ \percc$.
The progenitor model is tailored to be consistent with the averaged elemental abundances
of EMP stars presented by \citet{Cayrel04}.

The blue solid curve of Fig. \ref{fig:metal} show metal abundances.
We briefly summarize the different and common points of them with respect to those
in the present-day ISM.
In general, Pop III SN models show the alpha-element abundance enhancement relative to
the solar abundance ratio (${\rm [\alpha /Fe]} \simeq 0.4$).
In a model with a mass of $13 \ \Msun$, however, O and Mg are slightly underabundant relative to solar
(${\rm [O/Fe]}=-0.15$ and ${\rm [Mg/Fe]}=-0.23$, respectively).
Instead, Si and S are slightly overabundant (${\rm [Si/Fe]}=+0.11$ and 
${\rm [S/Fe]}=+0.17$, respectively).
We here focus on the progenitor model with a C-normal abundance (${\rm [C/Fe]} = 0.18$).
Note that there are another classification of EMP stars with
C-enhanced abundance ratio (${\rm [C/Fe]} > 2.3$; see Sec. \ref{sec:GA}).

The orange dotted curve of Fig. \ref{fig:metal} shows the condensation
efficiency of each element.
Figs. \ref{fig:dust} and \ref{fig:dist} show dust abundances and 
size distribution.
The mass ratio of dust relative to metal (condensation ratio) 
is only 9.3\%, which is less than that in the present-day ISM by a factor of five.
The major species are initially metallic silicon and amorphous carbon.
Silicate grains (forsterite and enstatite) are less abundant before grain 
growth occurs \citepalias[see][]{Chiaki15}. 
The size distribution is proportional to $r^{-3.5}$ \citep{Nozawa07} as 
in the present-day \citep{Pollack94}.

Throughout this paper, metallicity is normalized by the solar metallicity
$\Zsun = 0.01295$ and the solar elemental abundance ratio is taken from \citet{Asplund09}.


\begin{figure*}
  \vspace{2em}
 \includegraphics[width=\textwidth,natwidth=7.75cm,natheight=10cm]{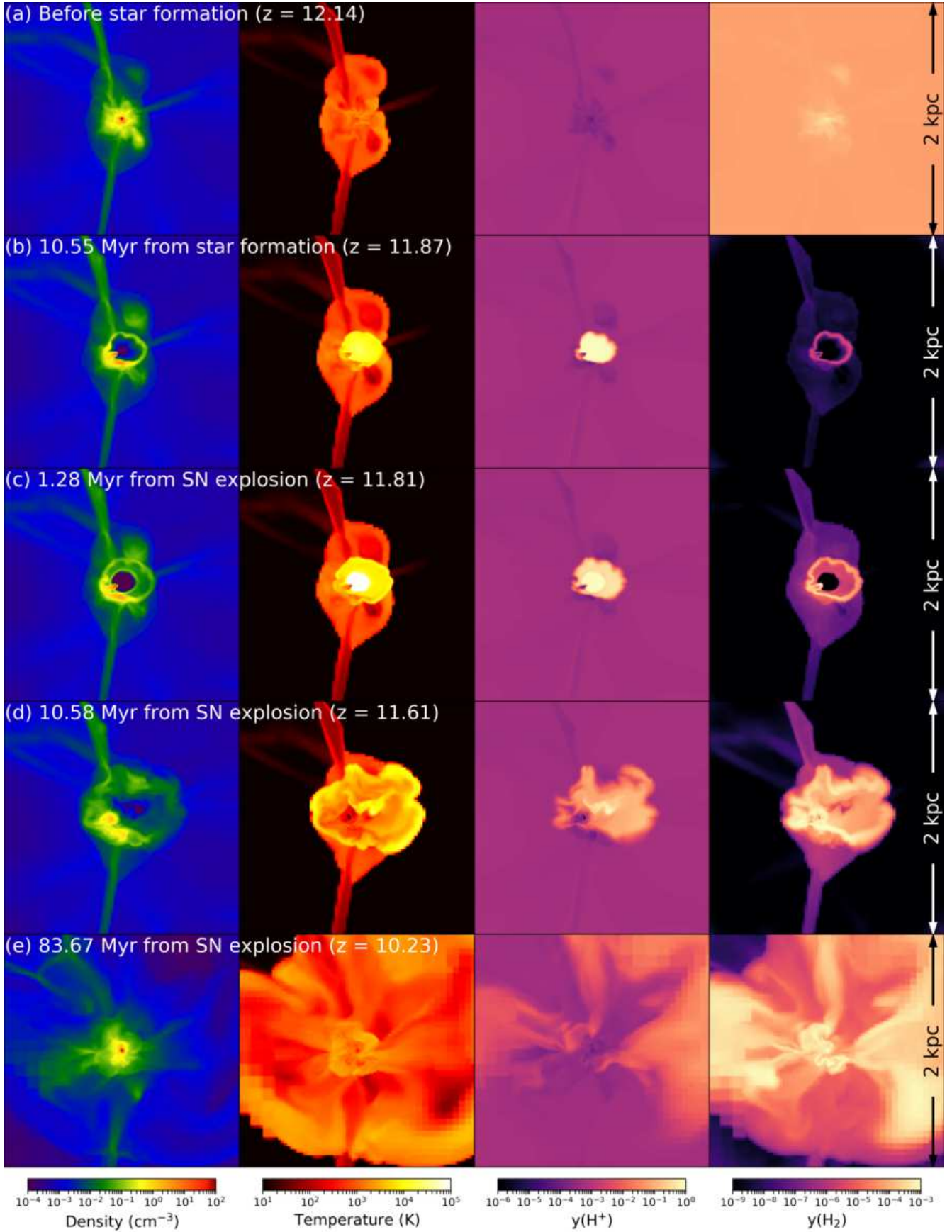}
\caption{
Slices of density, temperature, H$^+$ and H$_2$ abundances from left to right at the four stages
from top to bottom:
(a) collapse of a Pop III star-forming cloud, 
(b) 10.6 Myr after the star formation, 
(c) 1.3 Myr and
(d) 10.6 Myr after SN explosion, and
(e) when the halo recollapses.
}
\label{fig:output_time}
\end{figure*}

\begin{figure*}
\includegraphics[width=5.5cm,natwidth=7.25cm,natheight=6.5cm]{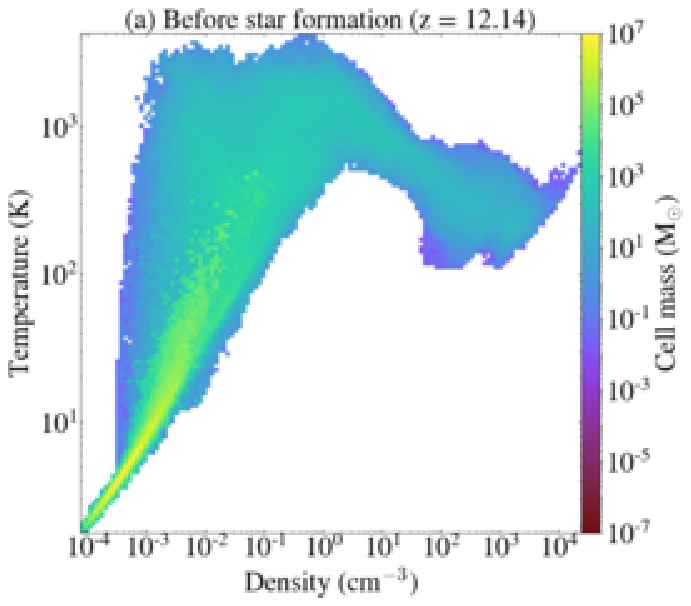}
\includegraphics[width=5.5cm,natwidth=7.25cm,natheight=6.5cm]{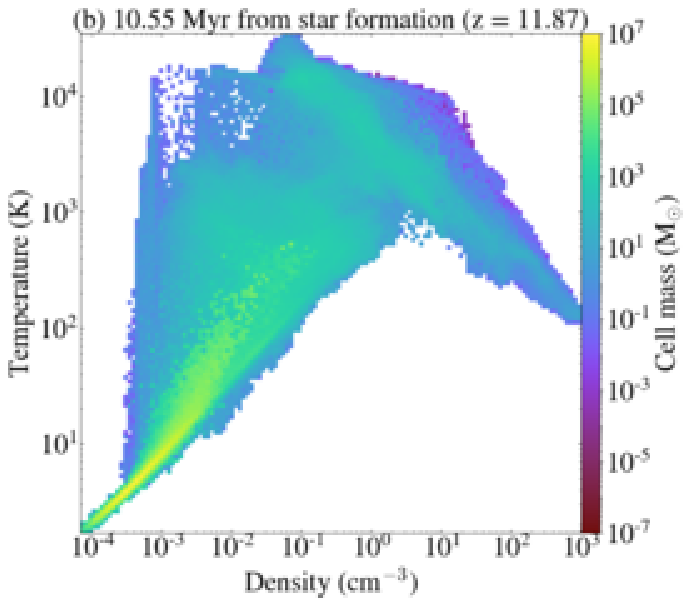}
\includegraphics[width=5.5cm,natwidth=7.25cm,natheight=6.5cm]{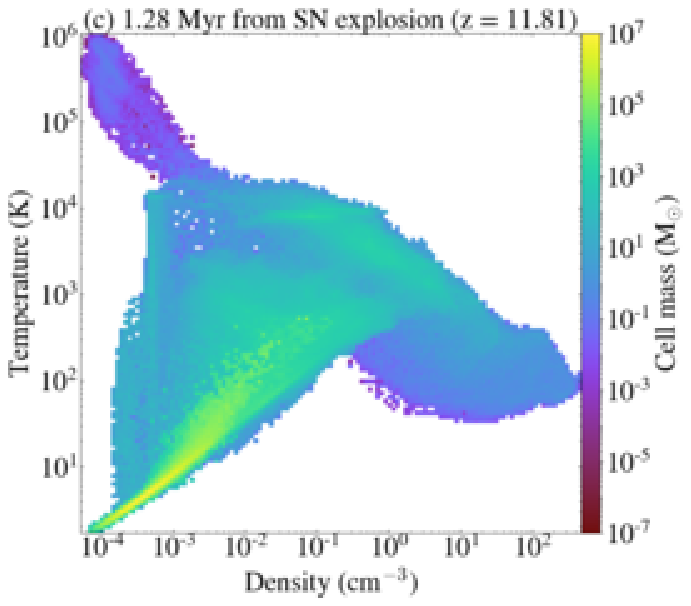}
\includegraphics[width=5.5cm,natwidth=7.25cm,natheight=6.5cm]{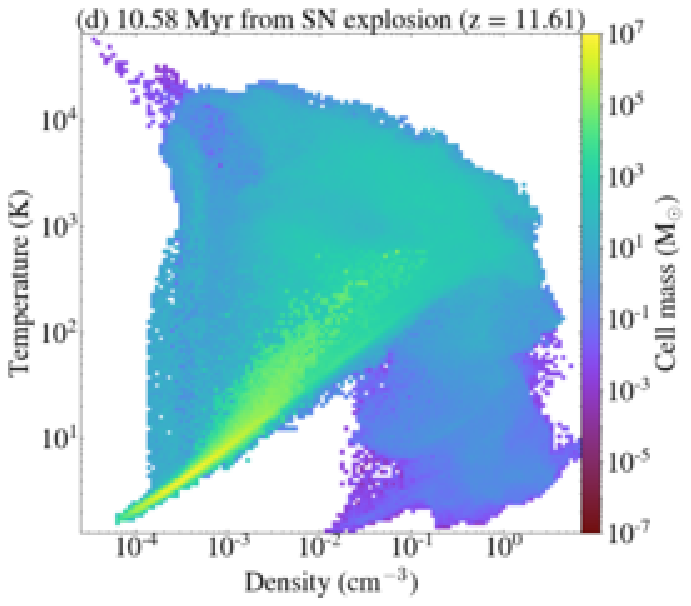}
\includegraphics[width=5.5cm,natwidth=7.25cm,natheight=6.5cm]{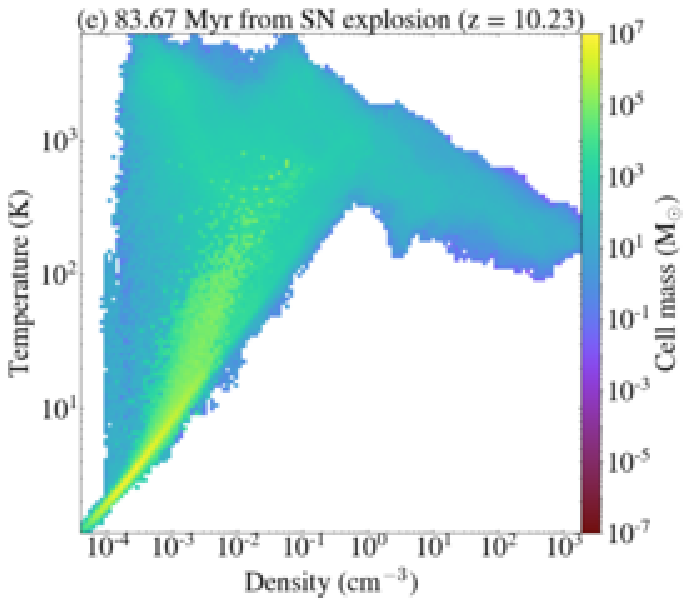}
\includegraphics[width=5.5cm,natwidth=7.25cm,natheight=6.5cm]{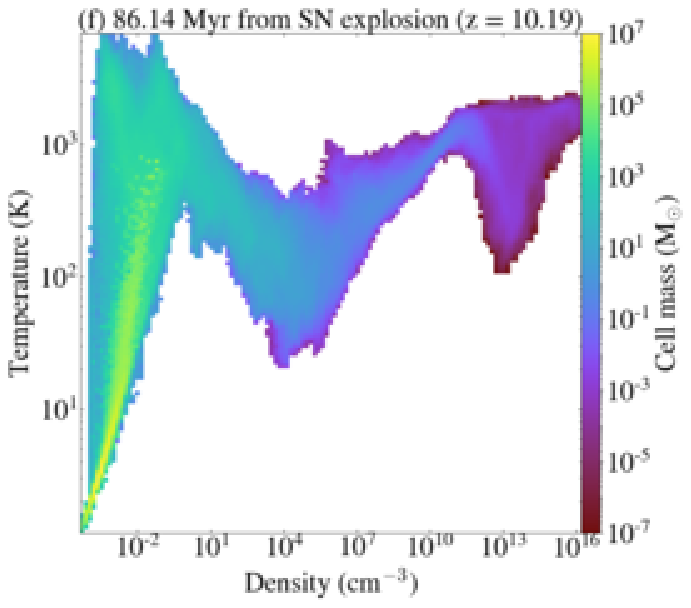}
\caption{
Two-dimensional histograms of density and temperature at the same time as in Fig. \ref{fig:output_time} (a--e) and 
when a protostar forms in the recollapsing cloud (f).
}
\label{fig:nT}
\end{figure*}

\begin{figure*}
\includegraphics[width=18cm,natwidth=12inch,natheight=3.5inch]{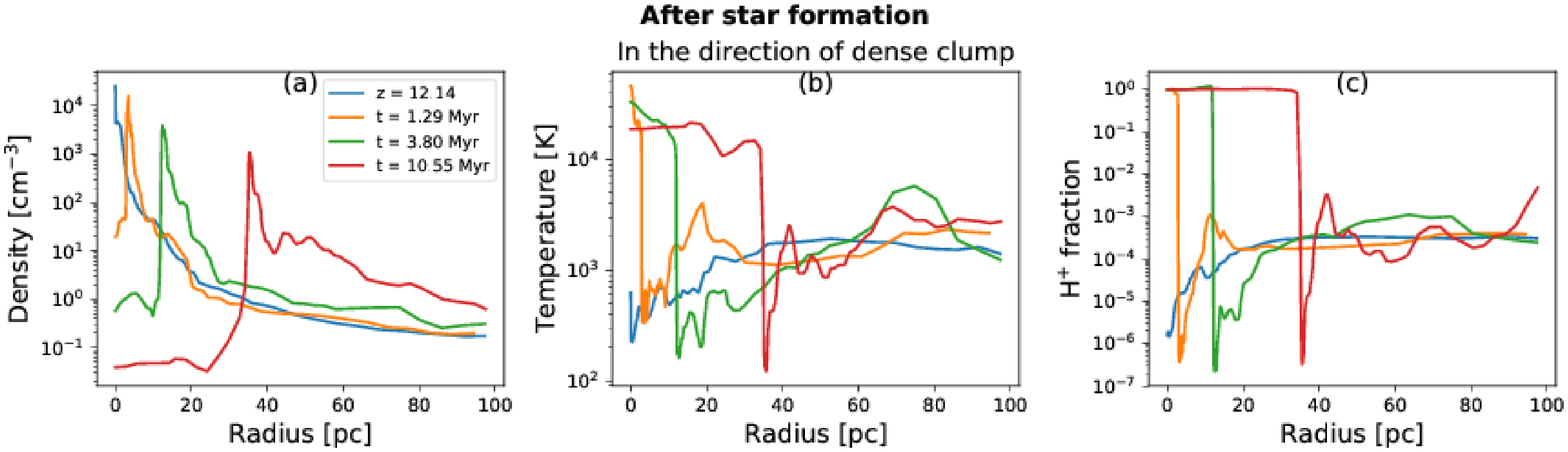}
\includegraphics[width=18cm,natwidth=12inch,natheight=3.5inch]{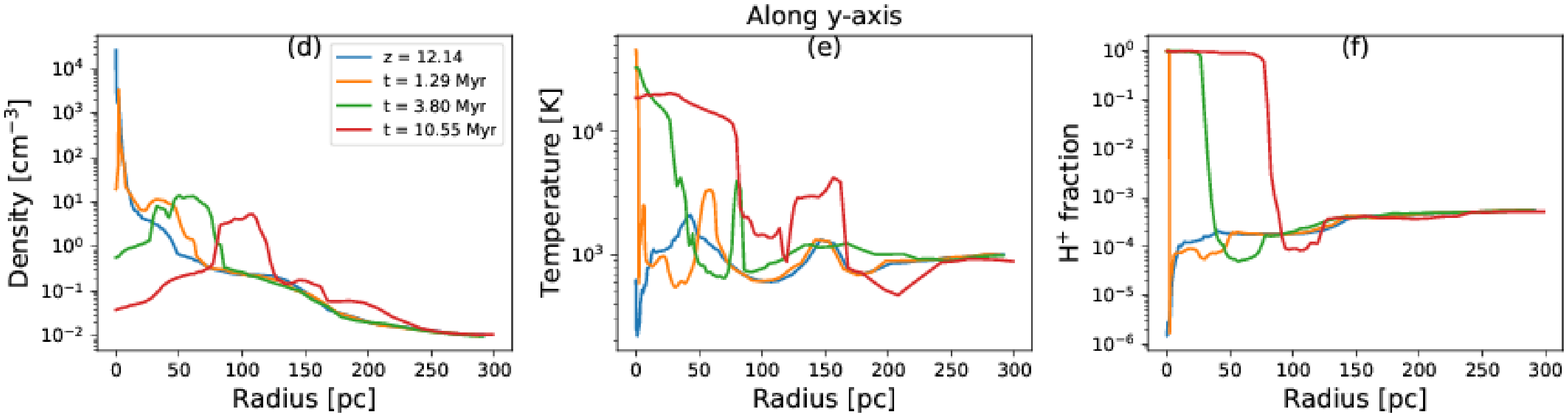}
\caption{
Density, temperature, and H$^+$ fraction
as a function of distance from the Pop III star in the direction of a dense clump
(top) and along the computational y-axis (bottom)
before (blue) and 1.3 (orange), 3.8 (green), and 10.6 Myr (red) 
after formation of the central Pop III star.
}
\label{fig:radial_profile_ion}
\end{figure*}

\begin{figure*}
\includegraphics[width=18cm,natwidth=12inch,natheight=3.5inch]{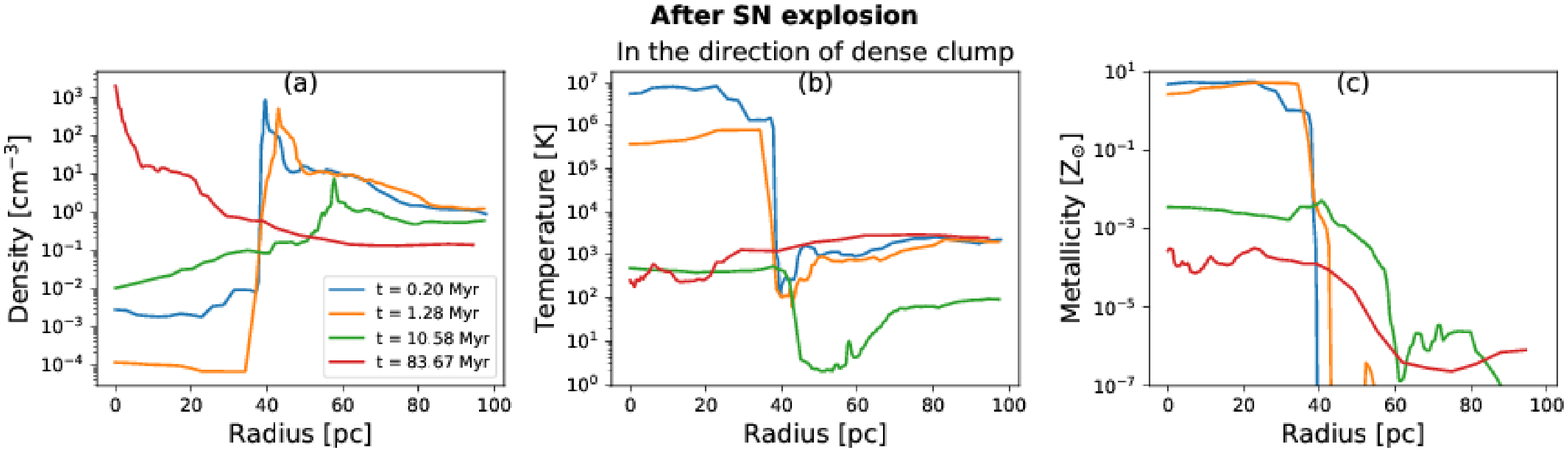}
\includegraphics[width=18cm,natwidth=12inch,natheight=3.5inch]{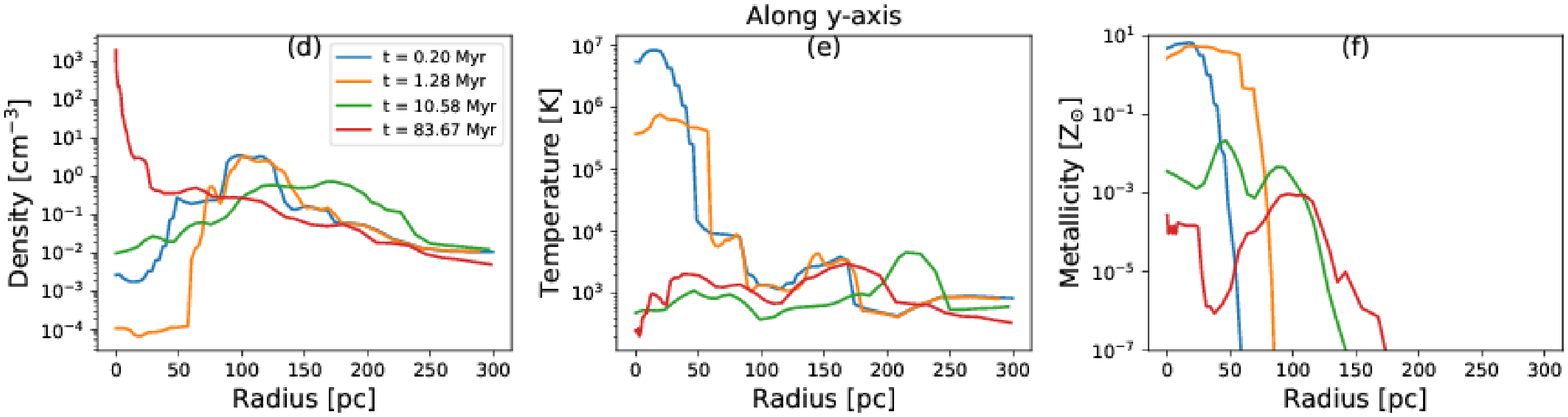}
\caption{
  Similar to Figure \ref{fig:radial_profile_ion} but
    focusing on the SN remnant.  Density, temperature, and metallicity
    as a function of distance from the stellar remnant at 0.2 (blue),
    1.3 (orange), 10.6 (green), and 24.5 Myr (red) after the SN
    explosion.  }
\label{fig:radial_profile_SN}
\end{figure*}

\section{Results}
\label{sec:results}

We follow the entire formation sequence from Pop III star formation to the near birth of a second-generation star
in a cosmological context.
Fig. \ref{fig:output_time} shows the slice of density, temperature, H$^+$, and
H$_2$ fractions at four characteristic epochs.
First, a MH with a mass of $1.77\E{6} \ \Msun$ virializes at the center of the 
simulation box at redshift $\zform = 12.1$, in which
the gas begins to collapse through hydrogen molecule cooling, forming a Pop III star with a mass of $13.5 \ \Msun$ (Fig. \ref{fig:output_time}a).
We follow the evolution of H {\sc ii} region
formed around the Pop III star by solving the radiative transfer equation.
During the lifetime of $\tlife = 11.8$ Myr, the ionization and D-type fronts expand to 40 pc,
smaller than the virial radius 287 pc of the host MH.
This is the condition for an ensuing SN shock to fall back to the host MH
\citep[][\citetalias{Chiaki18}]{Kitayama05}.

At the time $\tlife$, a SN explosion occurs, and metal and dust are released into the immediate environment (Fig. \ref{fig:output_time}b--d).
We add metal and dust to the ejecta with the mass, abundances of each gas-phase element and dust species, 
and dust size distribution that have been 
consistently calculated by hydrodynamics evolution of Pop III SN with progenitor mass
$13 \ \Msun$.
About 80 Myr after the SN, the enriched gas falls back into the host MH as predicted, and then
begin to recollapse (Fig. \ref{fig:output_time}e).
The metallicity of the recollapsing region is $2.6\E{-4} \ \Zsun$, corresponding to 
${\rm [Fe/H]}=-3.42$, indicating that a EMP star will form there.
We calculate all relevant chemical reactions and grain growth to investigate the thermal evolution 
and fragmentation properties of the low-metallicity collapsing cloud.
Finally, a protostellar core forms in the optically thick region.
Knotty filamentary structures appear around the protostar, where gas becomes unstable
through dust cooling.
This indicates that the cloud fragmentation occurs and a low-mass stellar cluster would form
if we extended the simulations.

\subsection{Pop III star formation and feedback}
Our zoom-in simulation focuses on a MH with mass $\Mhalo = 1.77\E{6} \ \Msun$
and radius $\Rhalo = 287$ pc that virializes at redshift $\zform = 12.1$.
Fig. \ref{fig:output_time}a shows the slices of density, temperature,
the abundances of $\rm H^{+}$ and $\rm H_2$ just before star formation, while
Fig. \ref{fig:nT}a shows gas temperature as a function of density at that same epoch.
The gas accreting into the halo collapses adiabatically until the density reaches $\nH \sim 1 \ \percc$.
Then radiative cooling by H$_2$ molecules becomes effective, and temperatures drop down to
$\sim 200$ K at a density $\sim 10^3 \ \percc$.
Cloud fragmentation does not occur at higher densities due to the lack of coolants, and
at these scales, the mass range of the forming stars is determined
\citep{Omukai00}.
The typical mass of the system is the order of the Jeans mass $\sim 10^2$--$10^3 \ \Msun$
at this scale \citep{Omukai00}.
Prior work have shown disk fragmentation in the accretion disk of the
subsequently formed protostar and a lower mass range in the eventually formed stellar cluster
\citep{Clark11, Greif11, Stacy12}.
In this MH, a Pop III star particle is created with a mass $\MPopIII = 13.5 \ \Msun$
that was randomly sampled from the assumed IMF (Eq. \ref{eq:PopIII_IMF}).

The star emits hydrogen ionizing photons and H$_2$ dissociation photons
at the rate of $\QH = 1.23\E{48} \ {\rm s}^{-1}$ and $\QHH = 1.72\E{48} \ {\rm s}^{-1}$,
respectively during its lifetime of $\tlife = 11.8$ Myr \citep{Schaerer02}.
Fig. \ref{fig:output_time}b
shows the snapshot at 10.6 Myr after the star formation.
A clump close to the Pop III star is partially disrupted but remains dense.
Also, the gas at the connecting point between the cloud and filament is compressed.
Fig. \ref{fig:radial_profile_ion} shows
  one-dimensional profiles of density, temperature,
and H$^+$ abundance during the Pop III main sequence from the star toward
the clump and filament.
A D-type shock front forms and eventually expands to 40 and 100 pc in the directions
of the clump and filament, respectively, at $\tlife$.
Within it hydrogen is fully ionized, and 
temperature of the H {\sc ii} region reaches 20,000 K from the balance between ionization
heating and Ly$\alpha$ cooling.
The large gas pressure causes the central density tone almost uniform with $0.03 \ \percc$.
Fig. \ref{fig:nT}b shows that
a diffuse ($\sim 0.1 \ \percc$) and warm ($\sim 20,000$ K) phase appears.
The density is smaller than that of previous result \citep[][\citetalias{Chiaki18}]{Jeon14}.
The lifetime of Pop III star with $13 \ \Msun$ in our case is slightly longer than 
with $15 \ \Msun$ in \citet{Jeon14}, and D-type front is expanded to larger radius,
resulting in more diffuse H {\sc ii} region.
Also, compared with the same progenitor mass model of our previous SPH simulation \citepalias{Chiaki18},
the spatial resolution in the central evacuated region around the Pop III star
is higher, better capturing the density structures in the H {\sc ii} region.
In \citetalias{Chiaki18} the mass resolution was $4.5 \ \Msun$, corresponding to the spatial resolution of 28 pc
for $\nH = 0.1 \ \percc$ while it is 3 pc in this work.

The SN explosion occurs afterwards, and the blastwave propagates through the H {\sc ii}
region.
In the direction of the dense clump (Fig. \ref{fig:radial_profile_SN}a--c), the SN shock
interacts with the clump of density $10^3 \ \percc$ that is 40 pc from
the Pop III stellar remnant.
The clump is partly disrupted and is forced outward from the SN shock pressure (green curve).
Its density decreases and the clump eventually falls back into the recollapsing region (red curve).
In the direction of filament (Fig. \ref{fig:radial_profile_SN}d--f), the SN shock merges with the 
D-type shock just after the transition from the Sedov-Taylor phase to pressure-driven 
snowplough phase (orange curve).
The shell becomes thicker (green curve) and eventually falls back (red curve).

Finally, 80 Myr after explosion, which is approximately the dynamical time of the central
MH, the diluted shell begins to fall back into
the central MH.
The clumps near the Pop III remnant and at the connecting point
between the halo and filament form the subsequent star-forming cloud
(Fig. \ref{fig:output_time}e)
as previous works have found \citep{Ritter16, Chiaki18}.
The density of the halo center increases again to $10^3 \ \percc$ at 84 Myr after the Pop III star forms, 
and a runaway collapse begins (red curve in Fig. \ref{fig:radial_profile_SN}), which is
exactly the scenario of internal enrichment.
At that time the dark matter halo mass increases to $2.99\E{6} \ \Msun$,
1.69 times larger than at the time of Pop III star formation.
Assuming that the turbulent velocity is the order of the circular velocity proportional to
$\Mhalo ^{1/2}$, it increases by 1.30 times, which
affects the fragmentation properties of the enriched cloud as well as SN driven shocks.

\begin{figure}
\includegraphics[width=\columnwidth,natwidth=15cm,natheight=12cm]{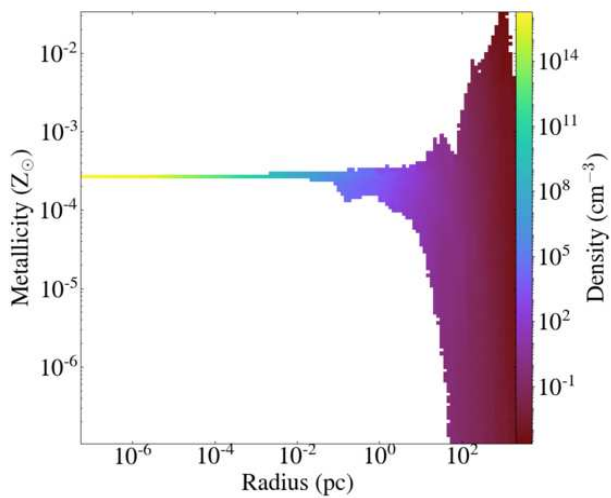}
\caption{
Metallicity distribution as a function of radius from the densest point
when the protostar is formed in the recollapsing region.
}
\label{fig:rZ}
\end{figure}

\begin{figure*}
 \includegraphics[width=18cm,natwidth=9.5cm,natheight=2.375cm]{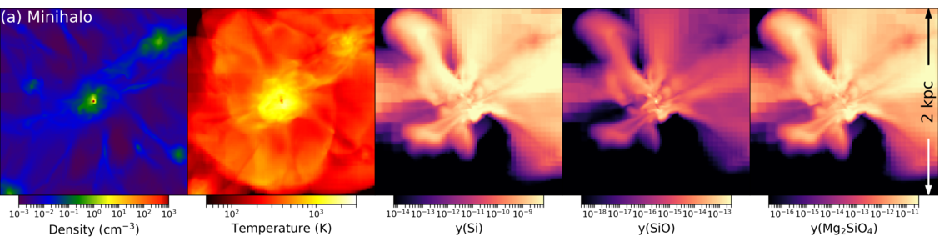}
 \includegraphics[width=18cm,natwidth=9.5cm,natheight=2.375cm]{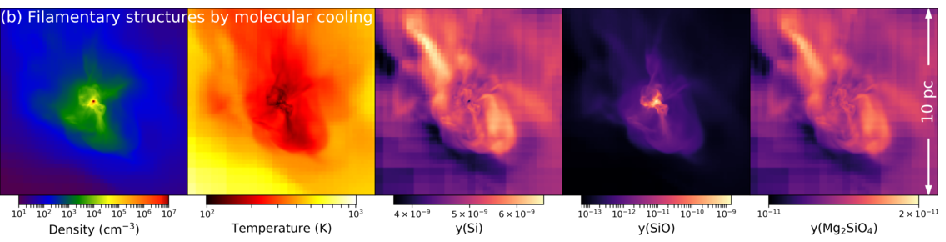}
 \includegraphics[width=18cm,natwidth=9.5cm,natheight=2.375cm]{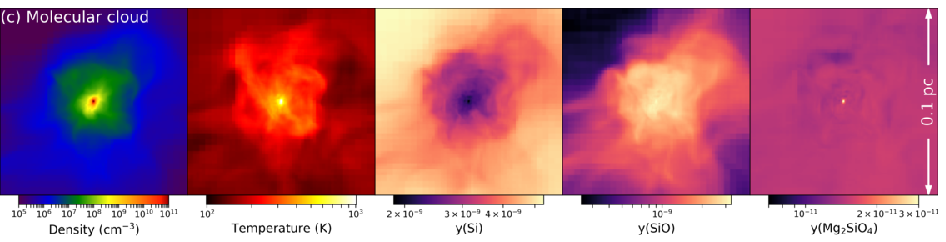}
 \includegraphics[width=18cm,natwidth=9.5cm,natheight=2.375cm]{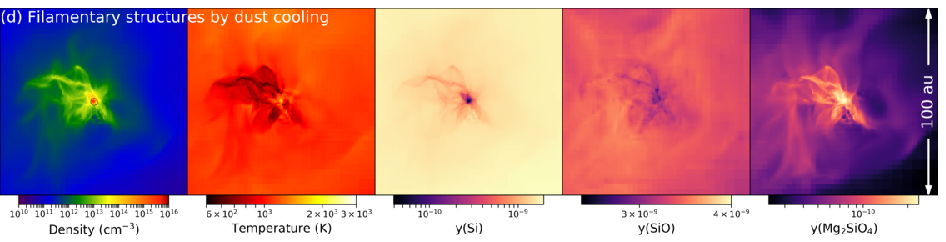}
\caption{
Density-weighted projections of density, temperature, Si, SiO, and forsterite (Mg$_2$SiO$_4$) absolute abundances from left to right.
The field of view progressively decreases from top to bottom, showing the morphology at halo scales
down to the protostar formation.
The black circle in the density map of panel (d) denotes the radius of the protostar.
}
\label{fig:output_zoom}
\end{figure*}

\begin{figure*}
\includegraphics[width=5.7cm,natwidth=4inch,natheight=3.2inch]{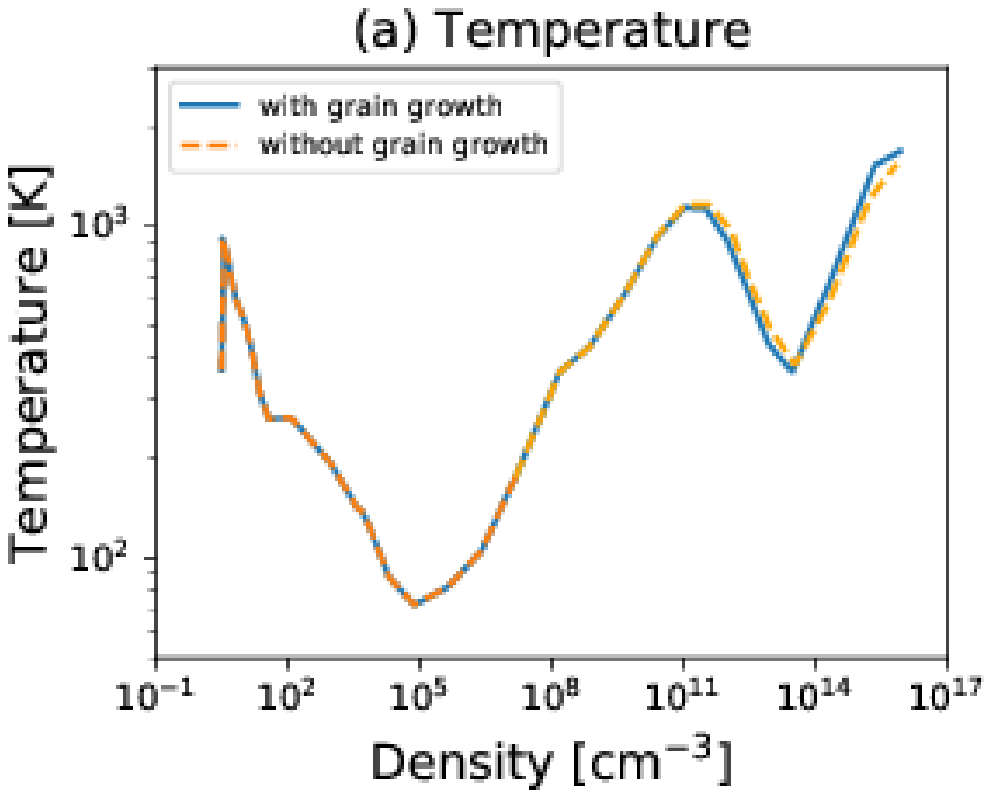}
\includegraphics[width=5.7cm,natwidth=4inch,natheight=3.2inch]{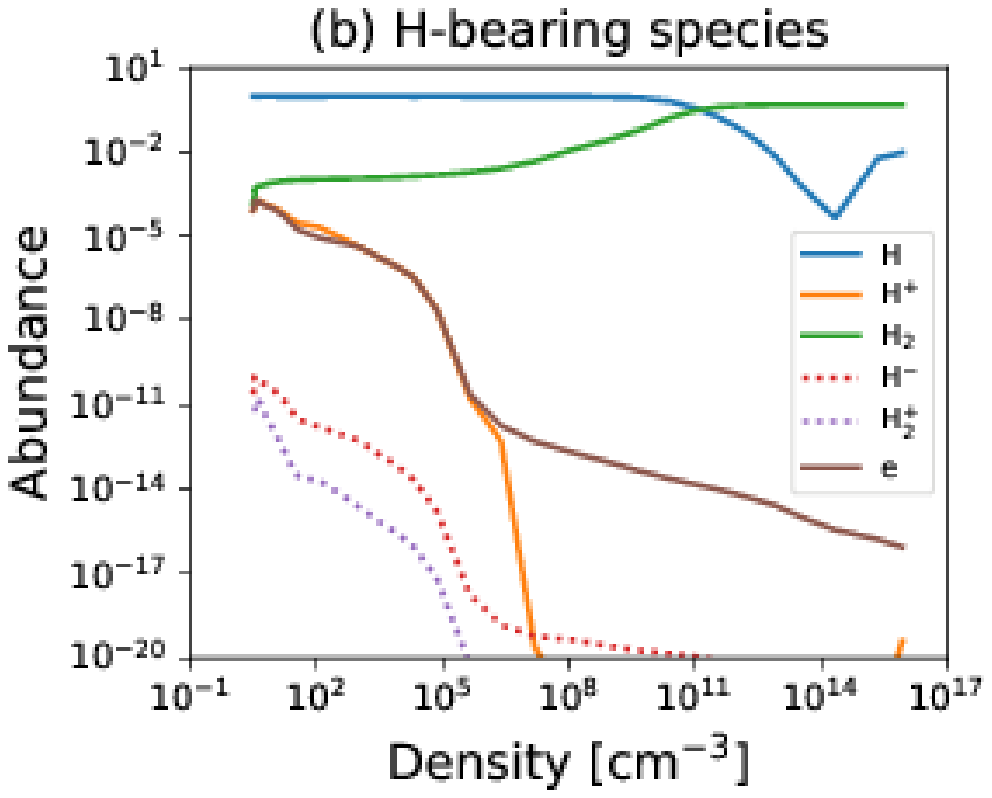}
\includegraphics[width=5.7cm,natwidth=4inch,natheight=3.2inch]{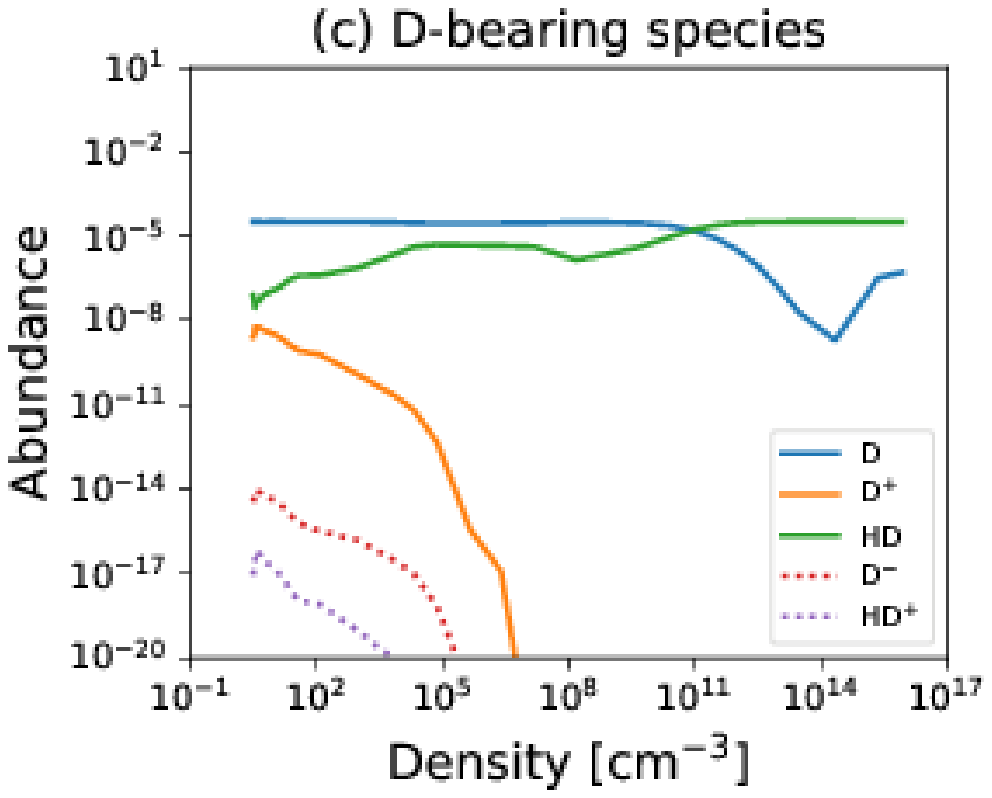}
\includegraphics[width=5.7cm,natwidth=4inch,natheight=3.2inch]{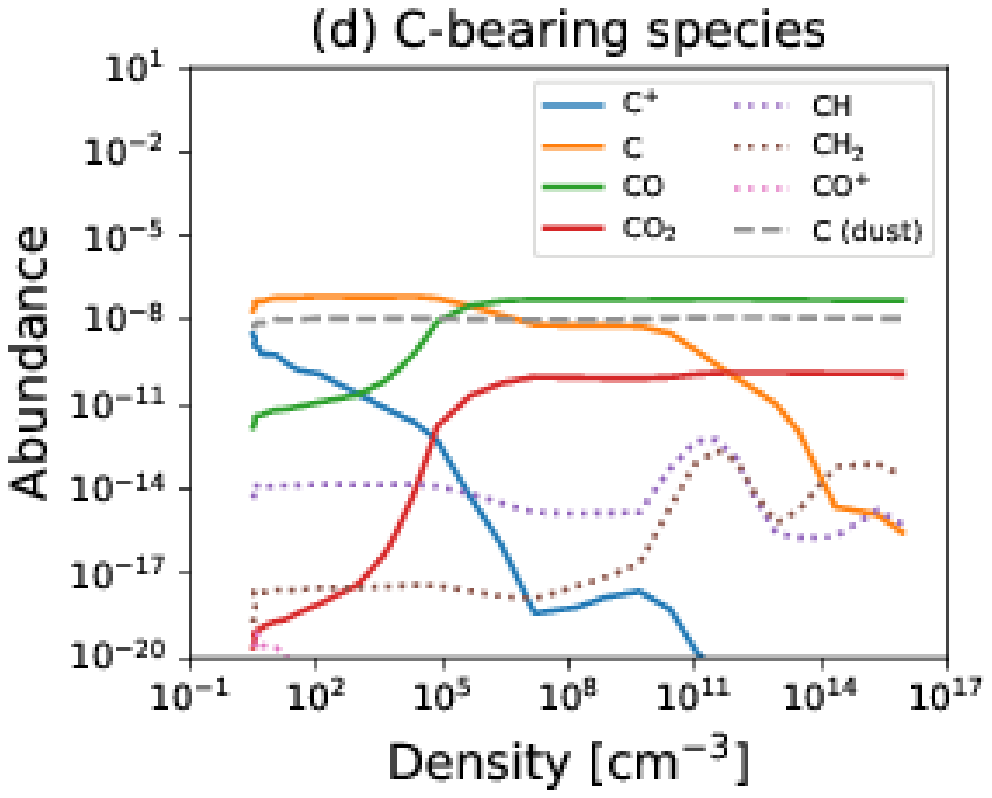}
\includegraphics[width=5.7cm,natwidth=4inch,natheight=3.2inch]{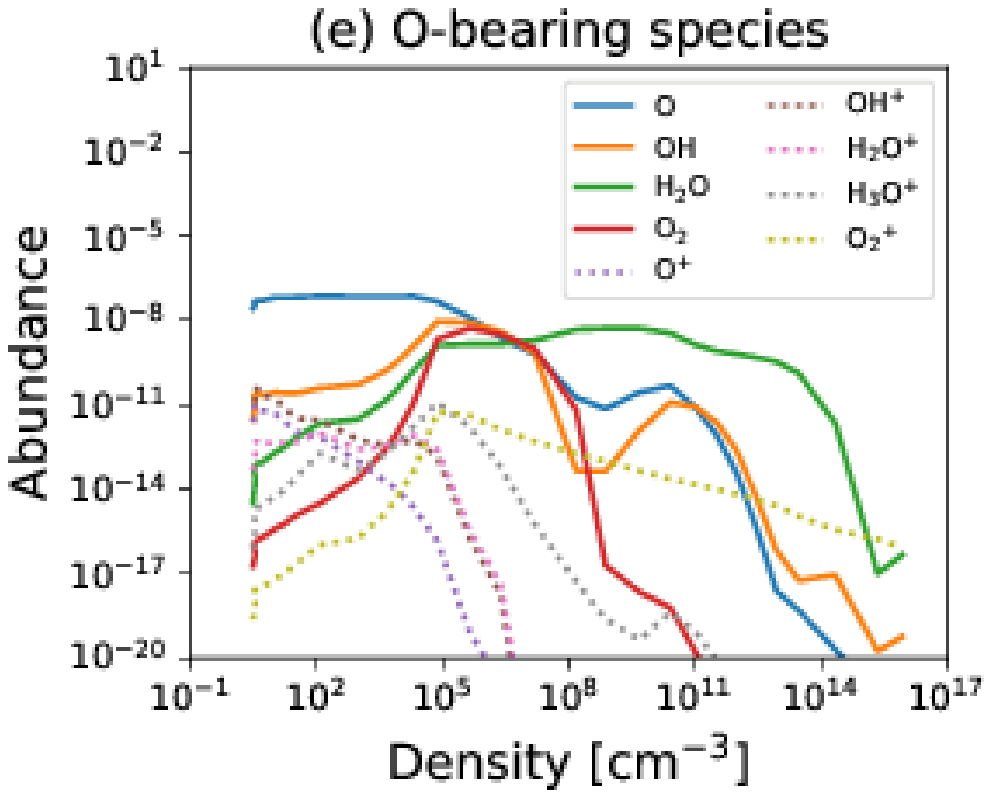}
\includegraphics[width=5.7cm,natwidth=4inch,natheight=3.2inch]{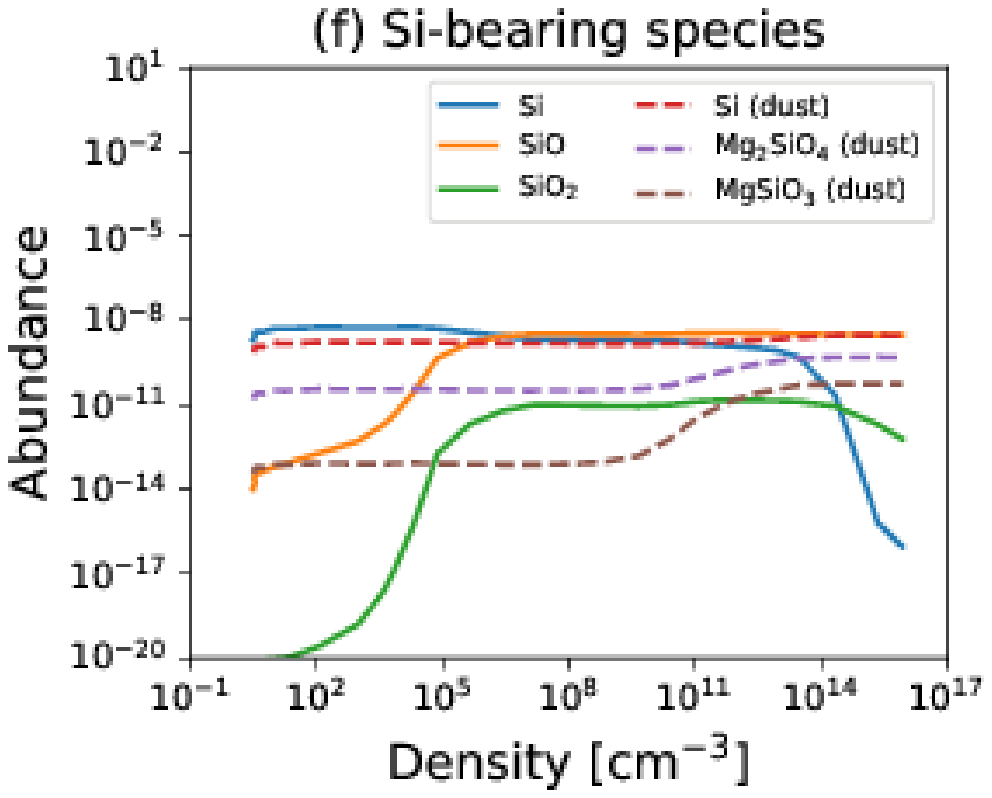}
\caption{
Temporal evolution of (a) temperature and number abundances of (b) hydrogen-, (c) deuterium-, (d) carbon-, 
(e) oxygen-, and (f) silicon-bearing species 
as a function of density in the recollapsing cloud core at each output time with its corresponding maximum density. 
We take the mass-weighted average of temperature and volume-weighted average of density of gas and the species in the core
in each snapshot.
The core is here defined as the region with densities 
$> n_{\rm H, max} /3$, where $n_{\rm H, max}$ is the maximum density.
In panel (a), we show the results of the simulation with (blue solid curve) and without (orange dashed curve) grain growth.
In panels (b)--(f), dotted and dashed curves show the abundances of gas-phase minor species and grain species,
respectively.
}
\label{fig:chemothermal_evolution}
\end{figure*}

\subsection{Metal-poor star formation}
\label{sec:thermal_evolution}
The metallicity in the recollapsing region is $Z_{\rm recol} = 2.6\E{-4} \ \Zsun$,
corresponding to iron abundance of ${\rm [Fe/H]} = -3.42$, indicating that
EMP stars will form in the cloud.
The abundances of the other elements are determined according to the 
elemental abundance ratio shown in Fig. \ref{fig:metal}.
In our previous study \citepalias{Chiaki18}, we terminated simulation when the maximum density 
of the recollapsing region becomes $1000 \ \percc$ to measure its metallicity
range.
In this work, we follow the subsequent chemical and thermal evolution to study the
fragmentation properties of the cloud until the maximum density reaches $10^{16} \ \percc$,
when the primordial component of gas becomes optically thick. 
Fig. \ref{fig:rZ} shows the metallicity distribution as a function of the distance
from the peak of the recollapsing cloud.
While there are both regions with large metallicity up to $10^{-2} \ \Zsun$ and with
pristine gas in the adjacent cosmological voids (radius $>100$ pc and density $< 0.1 \ \percc$),
the metallicity is uniform within $\sim 1$ pc
because only the inner part of fluid \citep{Smith15}, whose metallicity fluctuations are
progressively reduced, experiences a run-away collapse
\citep{Larson69, Penston69} faster than the metal mixing timescales.

In Fig. \ref{fig:output_zoom}, we show a sequence of zoom-in images focused on the recollapsing cloud.
Because of the adaptive nature of our simulation, we can resolve the cosmological volume with a side of $26.8$ kpc
(physical scale at redshift $z=10.2$) down to 0.010 au, corresponding to 33 levels of refinement.
Fig. \ref{fig:chemothermal_evolution} shows the evolution of temperature and chemical
abundances as a function of the central density.
Our detailed chemical network allows us to calculate the reactions including formation of metal molecules such as
CO, OH, and H$_2$O and growth of each grain species such as silicates for the first time.
The corresponding radiative cooling and chemical heating processes determine the evolution
of temperature, which affects the fragmentation properties of the gas cloud.

We here briefly summarize the chemo-thermal evolution of the recollapsing cloud
step-by-step.
At densities less than $10^3 \ \percc$, hydrogen molecules are formed via the H$^-$ 
process:
\begin{eqnarray}
\rm H + e^- &\to& \rm H^- + \gamma , \\
\rm H^- + H &\to& \rm H_2 + e^-
\end{eqnarray}
catalyzed by free electrons.
In gas with a metallicity of $2.6\E{-4} \ \Zsun$, H$_2$ formation on grain surfaces is
marginally dominant.
The H$_2$ fraction is $10^{-3}$ that is slightly higher than in the primordial gas 
(Fig. \ref{fig:chemothermal_evolution}b).
Their ro-vibrational cooling is dominant at that stage, and
temperature declines to 150 K.
At this low temperature, HD molecules form through the reactions
\begin{eqnarray}
\rm D + H_2 &\to& \rm HD + H , \\
\rm D^+ + H_2 &\to& \rm HD + H^+
\end{eqnarray}
(Fig. \ref{fig:chemothermal_evolution}c).
At $10^3 < \nH < 10^5 \ \percc$, due to HD cooling,
temperature further drops down to 70 K.
This is shown as the green-yellow region in the density map of Fig. \ref{fig:output_zoom}b,
where filamentary structures start to form.
Along the filaments, the temperature reaches a minimum value $\sim 100$ K, suggesting that
HD cooling induces the deformation of the gas cloud.

At this metallicity, metal fine-structure cooling is sub-dominant.
However, the transition discriminant introduced by \citet{Frebel13}
$D_{\rm trans} = \log (10^{\rm [C/H]} + 0.9\times 10^{\rm [O/H]}) = -3.09$
in our simulation, above the critical value of $-3.5$ where fine-structure
cooling is efficient.
We find that 
O atoms are converted into OH molecules primarily through the reaction
\begin{eqnarray}
\rm O + H &\to& \rm OH + \gamma 
\end{eqnarray}
at $\nH \sim 10^5 \ \percc$ (Fig. \ref{fig:chemothermal_evolution}e).
At the same time, 
C atoms are oxidized into CO molecules by
\begin{eqnarray}
\rm C + OH &\to& \rm CO + H
\end{eqnarray}
(Fig. \ref{fig:chemothermal_evolution}d).
At $10^5 < \nH < 10^8 \ \percc$, OH and CO molecules are the dominant coolants instead
of C and O atoms, and temperature remains around 100 K.

Because adiabatic compressional heating overcomes these cooling processes, the gas
slowly heats up.
The cloud becomes stable against deformation, and thus forms a hydrostatic spherical core
as shown by the red-colored region in density plot of Fig. \ref{fig:output_zoom}b.
Here a molecular cloud forms with a size of $\sim 0.1$ pc, which is depicted in Fig. \ref{fig:output_zoom}c
as the region where SiO molecules are abundant at a fraction $y({\rm SiO}) = 3\E{-9}$.
Si atoms are oxidized into SiO by the impact of OH and O$_2$ molecules
(Fig. \ref{fig:chemothermal_evolution}f).
At densities $10^8 < \nH < 10^{11} \ \percc$, gas heating along with H$_2$ formation 
through the three-body reactions
\begin{eqnarray}
\rm H + H + H &\to& \rm H_2 + H, \\
\rm H + H + H_2 &\to& \rm H_2 + H_2
\end{eqnarray}
become efficient, 
and deformation
of the molecular cloud is further suppressed.

Finally, at $10^{11} \ \percc$, grain growth occurs.
Fig. \ref{fig:chemothermal_evolution}f shows that forsterite (red dotted curve)
and enstatite (purple dotted curve) grows by accreting gas-phase Mg, SiO, and H$_2$O.
Fig. \ref{fig:chemothermal_evolution}e shows that H$_2$O molecules are totally depleted by grain growth
because the abundance of available H$_2$O molecules are less ($y({\rm H_2O})=5\E{-9}$) than 
that of Mg or SiO ($y({\rm Mg})=9\E{-9}$, $y({\rm SiO})=2\E{-8}$) just before grain growth.
Since oxygen is less abundant than carbon in the assumed progenitor model
(${\rm C/O} = 1.17$), most of oxygen has been depleted into CO molecules 
at $\nH \sim 10^5 \ \percc$ before the onset of H$_2$O molecule formation ($\sim 10^9 \ \percc$).
In Fig. \ref{fig:output_zoom}d, we can see that the
forsterite ($\Forsterite$) abundance is enhanced 
while SiO molecules are slightly depleted
along the dense filaments ($\nH \gtrsim 10^{13} \ \percc$).
A fraction of silicon has remained in its atomic form avoiding oxidization, but
Si atoms are eventually depleted into metallic silicon grains.
Fig. \ref{fig:output_zoom}d shows the absence of Si atoms in the very center of the
cloud ($\gtrsim 10^{14} \ \percc$).

The phase plot (Fig. \ref{fig:chemothermal_evolution}a) 
shows that dust cooling operates at densities 
$\nH \sim 10^{13} \ \percc$, and
the cloud again becomes unstable to deformation.
Fig. \ref{fig:output_zoom}d shows that filamentary structures (yellow region in density map)
appears along the cool region (black region in temperature map) with length of a few 10 au.
At density $\sim 10^{14} \ \percc$, dust cooling become ineffective at which point dust and gas are thermally
coupled and  the gas becomes optically thick through the dust opacity.
The gas evolves adiabatically, forming a hydrostatic core (red region in Fig \ref{fig:output_zoom}d), 
the so-called first core \citep{Larson69, Penston69}.
Hereafter, we call it a ``protostar''.\footnote{Strictly speaking, a protostar forms at 
a later stage. At $\nH \gtrsim 10^{16} \ \percc$, cooling by H$_2$ dissociation will make
the equation of state softer, after which a second collapse occurs.
After H$_2$ molecules are fully dissociated, the gas again collapses adiabatically.
The hydrostatic core embedded in the first core is called a protostar.}
Here we define the radius of the protostar to be where the thermal energy balances the gravitational energy.
The mass and radius of the protostar are $0.06 \ \Msun$ and 1.8 au, respectively.
These are consistent with the Jeans mass at density $10^{15} \ \percc$ and temperature 500 K,
where gas becomes optically thick due to absorption of continuum photons by dust grains.
The black circle in the density map of Fig. \ref{fig:output_zoom}d denotes the radius of
the protostar.
This result suggests that dust cooling induces the formation of the low-mass star.

We should note that the stellar mass is eventually
  determined when the accretion onto the protostar halts.
Also, we do not see any fragmentation that would be the precursor of further low-mass star formation.
It is still possible that the filaments around the protostar will fragment into low-mass protostars.
Several knotty structures appears along the filaments, and these are expected to be separately 
self-gravitating.
Because of computational limitations,
we can only follow the initial dynamics of the primal protostar.
Since the dynamical time is proportional to $\rho ^{-1/2}$, 
the necessary numerical timestep to follow the stable and accurate evolution of the dense
($10^{16} \ \percc$) primary protostar 
is too short to fully follow the accretion history of the less-dense ($10^{14} \ \percc$) filaments.
There are other approaches to extend the physical time for the simulation.
In the sink-particle method, dense regions that determine the computational time are artificially masked.
The gas within a given radius around a density peak is accreted onto an artificial particle mimicking a star
\citep{Stacy12, SafranekShrader14b}.
Another technique introduces a stiff equation of state to halt the further gas accretion into protostars
\citep[\citetalias{Chiaki16}; ][]{Hirano17}.
We will extend our simulation with one of these methods to follow the evolution of 
filamentary structures induced by dust cooling
to determine their fragmentation properties and final stellar masses in a future work.

\section{Discussion}
\label{sec:discussion}

\begin{figure}
\includegraphics[width=\columnwidth,natwidth=4inch,natheight=3.2inch]{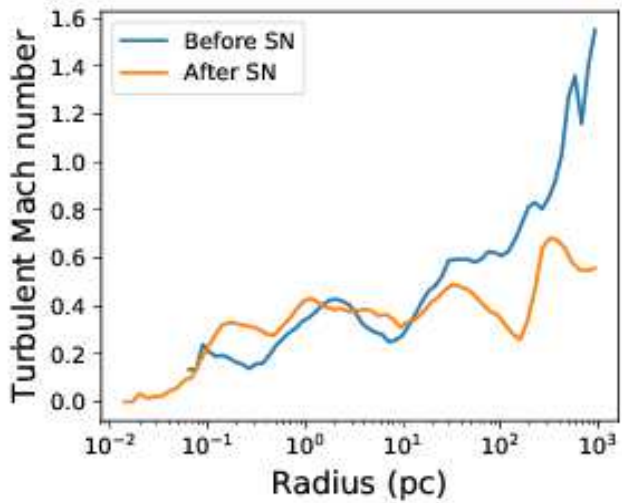}
\caption{
Radially averaged profiles of the turbulent Mach number, the ratio of
velocity dispersion and sound speed, in the Pop III star-forming cloud (blue curve)
and recollapsing cloud after the SN explosion (orange curve)
when the peak density is $2.6\E{4} \ \percc$ and $1.6\E{4} \ \percc$, respectively.
We subtract the bulk velocity calculated as the mass-weighted mean of velocity within 10 pc 
from the location of peak density.
}
\label{fig:radial_profile_vsig}
\end{figure}

\subsection{Effect of turbulence induced by SN shock}

In our previous work \citepalias{Chiaki16}, we followed the chemo-thermal evolution of
collapsing clouds from Pop III star-forming MHs with in the case of uniform metal distribution.
The conclusion was that the cloud fragmentation induced by dust cooling
halted in most cases because of preceding H$_2$ formation heating, which generally occurs
in low-metallicity clouds with $10^{-4}$--$10^{-3} \ \Zsun$.
We did not include any SN shock-driven turbulence.
\citet{Dopcke13} and \citet{SafranekShrader16} found that 
shocks added manually and driven by a halo merger, respectively,
drive density perturbations in the collapsing clouds.
These perturbations can become the seeds of fragments in the filamentary structures induced by
dust cooling.

In this work, we explicitly include the SN explosion and follow the subsequent shock.
Fig. \ref{fig:radial_profile_vsig} shows the turbulent Mach number
${\cal M}_{\rm turb} = \sigma _v / \cs$ as a function of distance 
from the density peak
for the Pop III star-forming cloud and the recollapsing cloud,
where $\sigma _v$ is velocity dispersion and $\cs$ is sound speed. 
The velocity dispersion is enhanced 
by gas accretion onto the halo and
by the SN-driven shock by a factor of at most two
within 10 pc from the density peak.
We measure the energy fraction of turbulence to the total energy in this region as
$\epsilon _{\rm turb} = \sigma _v^2 / (\bar{v}^2 + \bar{\cs} ^2)$ as
$0.13$ and $0.38$, respectively,
where $\bar{v}$ and $\bar{\cs}$ are mass weighted average of velocity magnitude and
sound speed in the core region with a radius of 10 pc.
The $\epsilon _{\rm turb}$ fraction increases in the latter by a factor of three.
As a result, multiple filamentary structures appear in our simulation with
SN shock-driven turbulence (Fig. \ref{fig:output_zoom}) while only a single
filament is seen in our previous work \citepalias[Fig. 5 in ][]{Chiaki16}.

\begin{figure}
\includegraphics[width=\columnwidth,natwidth=4.5cm,natheight=2.25cm]{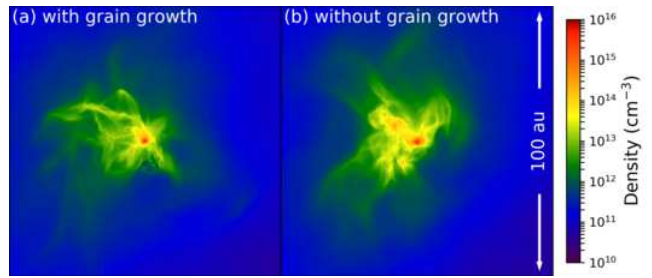}
\caption{
Density-weighted density projections at the maximum density $\nH = 10^{16} \ \percc$ in the runs
(a) with and (b) without grain growth. 
}
\label{fig:compare_slice}
\end{figure}

\begin{figure}
\includegraphics[width=\columnwidth,natwidth=14.5cm,natheight=13cm]{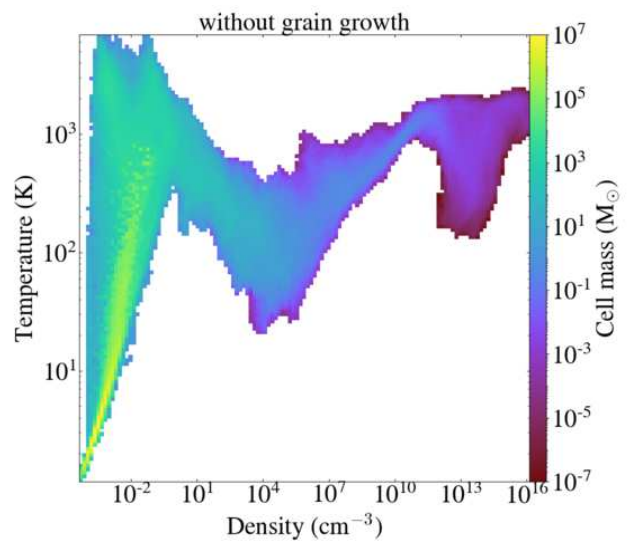}
\caption{
Same as Fig. \ref{fig:nT}f but for the simulation without grain growth.
}
\label{fig:nT_ng}
\end{figure}

\subsection{Effect of grain growth}

To study how grain growth affects the thermal evolution of metal-poor collapsing clouds,
we here also run a reference simulation without grain growth and
overplot the thermal evolution of the run in Fig. \ref{fig:chemothermal_evolution}a
with the orange dashed curve.
We find that the depth of the ``trough'' of temperature at densities $\nH = 10^{12}$--$10^{16} \ \percc$,
where dust cooling operates, is not so enhanced by grain growth.
Fig. \ref{fig:compare_slice} shows the snapshots of the central 100 au region
in the runs with and without grain growth.
These show the similar filamentary structures, and fragmentation notably appears even
in the latter case opposite to our expectation.
Fig. \ref{fig:nT_ng} shows the two-dimensional histogram of the density and temperature
in the run without grain growth when the maximum of gas density reaches $10^{16} \ \percc$.
We find that no significant difference is seen compared with Fig. \ref{fig:nT}f for the run with
grain growth.
The difference of the fragmentation properties could stem from the minute differences in the chemo-thermal 
evolution during the collapse, altering its timing and filamentary structures.

In the simulation with a metal and dust model with a progenitor mass $\MPopIII = 13 \ \Msun$,
the growth of silicates, which are the major species for gas cooling,
is suppressed because oxygen has been depleted into CO molecules at $\nH \sim 10^5 \ \percc$.
If ${\rm C/O} < 1$, a fraction of oxygen is reduced into H$_2$O molecules, which are accreted onto
silicates without being depleted into CO molecules.
This scenario occurs for progenitor masses of $\MPopIII > 20 \ \Msun$ as we show that silicates grow sufficiently
to affect the thermal evolution in \citetalias{Chiaki16}.

\begin{figure*}
\includegraphics[width=8.5cm,natwidth=4cm,natheight=2.5cm]{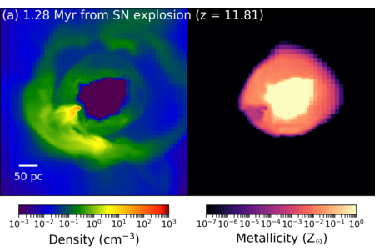}
\includegraphics[width=8.5cm,natwidth=4cm,natheight=2.5cm]{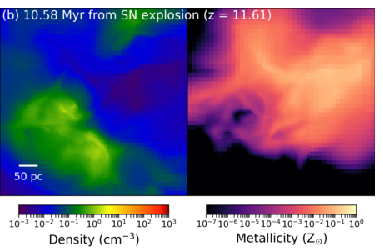}
\includegraphics[width=8.5cm,natwidth=4cm,natheight=2.5cm]{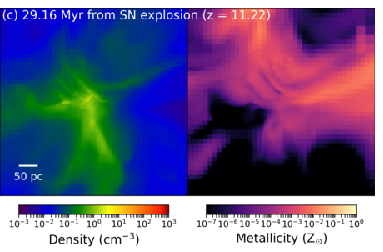}
\includegraphics[width=8.5cm,natwidth=4cm,natheight=2.5cm]{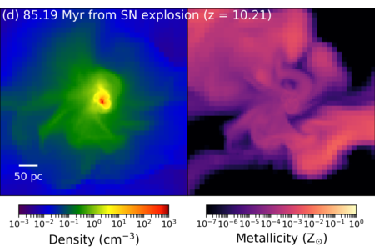}
\caption{
Density-weighted projections of density and metallicity with a field of view of $500$ pc and a depth of $100$ pc
at the following epochs: (a) 1.3, (b) 10.6, (c) 29.2, and (d) 85.2 Myr after supernova explosion
centered at the centroid of the MH.
}
\label{fig:time_series_SN}
\end{figure*}

\subsection{Simplified enrichment model}

We can further understand the chemical enrichment through a simplified model from a single event. 
The recollapsing cloud in the simulation has a metallicity of $2.6\E{-4} \ \Zsun$. We define the cloud as a sphere 
with a mass $M_{\rm cloud} = 2\E{3} \ \Msun$ of the Jeans mass at a density of $10^3 \ \percc$
and temperature of 100 K that has been uniformly enriched in the simulation 
within a radius of $1$ pc (see Fig. \ref{fig:rZ}). 
We can estimate the average metallicity of the cloud 
by assuming that all of the ejecta metal mass $M_{\rm met} = 1 \ \Msun$ is reincorporated into the cloud, 
giving a metallicity
\begin{equation}
Z_{\rm pred} = \frac{M_{\rm met}}{M_{\rm cloud}} = 4\E{-2} \ \Zsun.
\end{equation}
However, in reality, most of the metal escapes from the halo along with the shocked gas
into the adjacent voids.
The rest recollapses into the cloud, being blocked by the cosmic filaments and a
neighboring gas clump.

Fig. \ref{fig:time_series_SN} shows the process of metal dispersion along
with the SN shock propagation and fall-back into the central MH.
At 1.3 Myr after explosion (panel a), the SN shock with radius 40 pc is still inside the D-type shock
with radius 100 pc, and begins to interact with the neighboring clump.
The initial enriched gas does not mix with the interior gas 
at regions with densities $\gtrsim 3 \ \percc$ (yellow-orange colored region) 
and passes around the clump.
The partially destroyed clump and two connecting regions between
the MH and cosmic filaments remains dense,
and the SN shock stalls out
at 10.6 Myr after the SN (panel b).
The shock does not escape the halo's gravitational potential well, and 
these three regions accrete back into the
MH (panel c). 
The metals are then incorporated into the recollapsing cloud
and mixed by the turbulence driven by the SN and virialization (panel d).

With this picture, we can estimate a fraction $f_{\rm fb}$ of metals which fall-back
as \citet{Ritter15} did.
The fraction of metals that is blocked by the clump is
the ratio of solid angles of neighboring clumps with radius $r$ at a distance $R$ from the
SN is $\pi r^2 / 4\pi R^2$.
The radius of the region with densities $\nH \gtrsim 3 \ \percc$ is $r = 5$ pc
and the distance is $R = 50$ pc (Fig. \ref{fig:time_series_SN}b).
There are $N_{\rm clump}=3$ of clumps that interrupt the propagation of metals.
We can thus estimate the metallicity $Z_{\rm recol}$ in the recollapsing region as
\begin{equation}
Z_{\rm recol} = \frac{f_{\rm fb} M_{\rm met}}{M_{\rm cloud}} 
= N_{\rm clump} \frac{\pi r^2}{4\pi R^2} \frac{M_{\rm met}}{M_{\rm cloud}}
= 3\E{-4} \ \Zsun,
\end{equation}
which is consistent with our simulation result.

\subsection{Minihalo and Pop III star mass}

In this work, we investigate the metal enrichment in
  the case of a particular scenario: an initial MH mass $\Mhalo = 1.77\E{6} \ \Msun$ and 
Pop III stellar mass $\MPopIII = 13.5 \ \Msun$.  The subsequent
dynamics and metal enrichment heavily depend on these two
characteristic properties, where the ratio of the halo binding energy
and the radiation/explosion energies is the key quantity.
However, we can categorize this spectrum of initial parameters and
associated outcomes into two scenarios:  internal enrichment and external enrichment.
In our previous work \citepalias{Chiaki18}, we found that the external enrichment occurs
when ionizing region expands beyond the halo virial radius with MH masses below the critical 
mass:
\begin{eqnarray}
M_{\rm halo, cr}^{\rm ion} &=&
6\E{5} \ \Msun
\left\{
\left( \frac{\MPopIII}{25 \ \Msun} \right) 
\right. \nonumber \\ && \times \left.
\left[
1-\exp
\left( -\frac{\MPopIII}{25 \ \Msun} \right) 
\right]
\right\} ^{3/4}.
\label{eq:MHion_fit}
\end{eqnarray}
In this case, because the SN shell expands through the low-density region ($\lesssim 10 \ \percc$),
it does not sufficiently lose thermal energy due to the low cooling rate (proportional to density squared) and continues to expand.

In the simulated mass range of $\Mhalo = 2\E{5}-1\E{6} \ \Msun$ \citep{Hirano14}, the main enrichment process is internal
enrichment for core-collapse SNe ($\MPopIII = 8-40 \ \Msun$).
In the most energetic case, metals can escape from the host MH and external enrichment occurs for pair-instability
supernovae \citep[$\MPopIII = 140$--$260 \ \Msun$;][]{Heger02}. 
Therefore, to fully investigate the metal distribution function in the
Universe, it is required to follow the
metal dispersion from a multitude of host halos and Pop III stars with
a wider sampling of their initial masses and thus stellar endpoints.
However by focusing on individual cases, we can further understand
the detailed dynamics and mixing during the transition from Pop III
to metal-enriched star formation.

\subsection{Isolation of minihaloes}
Our simulation focuses on an isolated MH in a
  cosmological setting.  Particularly in large-scale overdense
  regions, MHs can be clustered enough so that their SN remnants will
  overlap.  The mixing of nucleosynthetic products from multiple
  events will partially wash out any unique feature from each
  explosion \citep{deBennassuti14, deBennassuti17, Hartwig18}.

It has been shown that metal-free stars generally form
  in halos of mass $10^6~\Msun$ \citep{Bromm13}.  For a large-scale
  region with a typical overdensity, such as the one that eventually
  forms the Milky Way, the average MH separation can be calculated
  with (elliposidal) Press-Schetcher formalism \citep{Press74,
    Sheth01} to be $200 \left( \Mhalo / 10^6 \ \Msun \right) ^{0.5}$
  comoving kpc at redshift $z=15$.  This is comparable with the
  simulation side length (300 comoving kpc), justifying our choice of
  considering a single enrichment event.  We should note that,
  however, some fraction of MHs will be clustered in the Milky Way
  progenitor environment.  In such regions, EMP stars should be
  enriched by multiple progenitors.  This suggests that EMP stars,
  especially for C-normal ones, are not the only second-generation
  stars which we investigate in this work.  To capture the statistical
  feature of first stars and EMP stars, simulations in a larger
  cosmological volume are required with a special focus on progenitors
  of Milky Way like galaxies \citep{Tumlinson07, Salvadori07,
    Sarmento17}.

\subsection{Shell instability}
When the SN shock interacts with the D-type shock, there is a possibility that
the shell becomes sufficiently thick to become unstable against fragmentation.
In the dense clump neighboring the SN, its density is $10^3 \ \percc$ when the shocks interact
(Fig. \ref{fig:radial_profile_SN}a).
However, the clump is partly desrupted by the pressure of SN shock before it collapses.
Also, in the direction of cosmic filament (Fig. \ref{fig:radial_profile_SN}d), the gas falling into the MH collide with the SN shock, forming a dense shell.
The intability criterion of an expanding shell with velocity $V_{\rm sh}$ at distance $R_{\rm sh}$ 
from the SN can be written as
\begin{equation}
\frac{V_{\rm sh}}{R_{\rm sh}} < \frac{\pi G \Sigma _{\rm sh}}{\sqrt{8}c_{\rm eff}},
\end{equation}
where $\Sigma _{\rm sh}$, $c_{\rm eff}$ are the surface density and effective sound speed of the shell
\citep[see][]{Chiaki13}.
For the dense shell with $V_{\rm sh} = 20$ km/s, $R_{\rm sh} = 100$ pc, $\Sigma _{\rm sh} = 8.1\E{-4} \ {\rm g/cm^2}$,
and $c_{\rm eff} = 2.2$ km/s, the left and right hand sides are
$0.20$ and $0.0085 \ {\rm Myr^{-1}}$, respectively, showing that the shell is also marginally stable.

\section{Comparison with observed EMP stars}
\label{sec:GA}

Recent observational campaigns have covered a large sample of EMP stars
\citep{Yoon18, Placco18}. 
Although we restrict our focus on a single metal-poor 
star-forming cloud enriched by a single Pop III SN,
this study gives a general insight for the chemo-thermal evolution of metal-poor clouds
and their prior enrichment.
We expect that the chemical/thermal processes such as
metal molecular cooling, H$_2$ formation heating, and dust thermal emission cooling 
as well as grain growth will be commonly crucial for cloud formation and collapse
because the thermal evolution of those clouds is nearly insensitive to their
initial conditions \citep[see discussion around Fig. 2 in][]{Omukai00}.
Although our study shows that cloud fragmentation is hardly seen until we terminate the simulation,
it is possible that knotty filamentary structure will fragment in
later phases.
However, these filaments can still accrete onto the central initial protostar, and if this accretion flow
remains intact until the protostellar radiation starts to sweep up the ambient gas,
low-mass EMP star formation will be suppressed \citep{Hosokawa11}.
The simple chemical evolution model of \citet{Hartwick76} suggests 
that the metallicity distribution of stars 
is proportional to $Z^{-1}$, while observations have shown that the number
of stars deviates below this prediction \citep{Frebel13},
which is consistent with our results in that low-mass EMP stars are rarely formed.

Our study, where we follow the advection of each element and chemical reactions,
can also be applied to the explanation of the existence of different
classifications among EMP stars.
Observations have revealed that EMP stars are classified into two major groups,
C-normal and C-enhanced populations \citep{Beers05}.
The vast majority of EMP stars with $\rm [Fe/H] < -4.5$ have carbon enhancement
\citep{Placco18, Yoon18}.
Although we here restrict our focus on the C-normal EMP stars,
the evolution of C-enhanced EMP star-forming clouds is also followed with 
progenitor models consistent with their elemental abundance ratio such as
faint SN \citep{Umeda03, Marassi14, Marassi15, Chiaki17}.

Lastly, we restrict our focus to a single EMP star-forming cloud enriched by
a single Pop III SN.
To study the nucleosynthesis and mass distribution of Pop III stars and the first metal
enrichment process indirectly from the metallicity and elemental abundance ratios
of EMP stars, it is preferable to isolate stars that are enriched by a single event 
\citep{Ishigaki18, Hartwig18}.
However, some EMP stars could be enriched by multiple progenitors \citep{deBennassuti14, 
deBennassuti17, Hartwig18}.
In our simulation, the recovery time during which the ejected gas returns to the
MH and ensuing star formation occurs is $\sim 80$ Myr.
In the case where other explosions in neighboring halos occur during this delay time,
the stars could be enriched by multiple progenitors, and the metallicity and elemental abundances
will be some combination of the progenitors' explosion characteristics.
Furthermore, a single halo might host multiple Pop III stars caused by fragmentation
\citep{Turk09, Clark11, Greif11, Stacy12, Susa14}.
To statistically compare with the large observational sample of EMP stars,
we plan to extend our simulation temporally and spatially to follow the effects of
multiple Pop III stars and formation of a population of EMP stars in future work.

\section{Summary and conclusion}
\label{sec:conclusion}

We follow the entire formation sequence of an
EMP star in a cosmological context.
In this simulation, we for the first time solve all relevant chemical 
reactions and grain growth in a low-metallicity molecular cloud,
considering nucleosynthesis and nucleation models of Pop III SN in a consistent manner.
In the targeted minihalo, metal-free gas contracts through hydrogen molecule cooling, where
a Pop III star forms with a mass $\MPopIII = 13.5 \ \Msun$.
We compute the radiation hydrodynamics of the system and follow the formation of an ionized region
around the Pop III star.
At the end of its lifetime, the ionization and D-type shock fronts reaches a radius of 40 pc, still contained within the MH that has a virial radius of 287 pc.  This containment is one of the conditions for
the self-enrichment of host MHs \citepalias{Chiaki18}.
This Pop III star explodes as a normal core-collapse SN with explosion energy $10^{51}$ erg.
In the ejecta, we uniformly add metal and dust models calculated for a Pop III SN with mass $13 \ \Msun$.
The ejected gas falls back into the central MH 80 Myr after the SN explosion.
In the recollapsing cloud, the metallicity is $2.6\E{-4} \ \Zsun$ nearly uniform 
within 1 pc.
We follow the further evolution of recollapsing cloud until the first protostellar core forms, at which time
the spatial resolution reaches a value of 0.01 au, sufficient to resolve the protostellar core.
We finally find that gas around the protostar deforms to 
multiple knotty filamentary structures by dust thermal emission cooling.
Although we do not follow the further evolution of these filaments due to the computational limits,
the filaments will fragment into a multiple protostellar system.

We show that a single SN event can reproduce a collapsing cloud with a metallicity
in a range of observed EMP stars through the internal enrichment process.
The mass of the primary protostar is $0.06 \ \Msun$ when we terminate the
simulation.
The protostar will continue to accrete the ambient gas, and the stellar mass will ultimately be determined when accretion is halted.
If this mass is less than $0.8 \ \Msun$, it will survive until the present-day.  If such an object formed in the Milky Way 
progenitor halos, it would be observed as an EMP star in the Galactic
halo or neighboring dwarf galaxies.
This work is the very first step to seek the chemical enrichment and star formation in the early universe.
By applying our strategy and results to larger simulations both in time and volume,
we will continue to clarify their entire formation sequence and connection to EMPs in the local universe, strengthening the insights gained from near-field cosmology.

\section*{ACKNOWLEDGMENTS}

We thank the anonymous referee for their constructive comments, improving our paper.
We also thank T. Nozawa, who kindly give us the SN models, and V. Bromm and
T. Hartwig for fruitful discussions.
We also thank B. Smith, who helped us with the {\tt yt} clump finding method.
GC is supported by Research
Fellowships of the Japan Society for the Promotion of Science (JSPS)
for Young Scientists.  JHW acknowledges support from NSF grants
AST-1614333 and OAC-1835213, Hubble Theory grant HST-AR-14326, and NASA grant
NNX-17AG23G.  The numerical simulations in this work are carried out
on PACE cluster in the Georgia Institute of Technology.  The freely
available plotting library {\sc matplotlib} \citep{matplotlib} was
used to construct numerous plots within this paper. Computations and
analysis described in this work were performed using the
publicly-available {\sc Enzo} and {\tt yt} codes, which is the product of a
collaborative effort of many independent scientists from numerous
institutions around the world.


\appendix
\section{Dust cooling rates with grain growth}
\label{sec:rates_dust}
Dust grains have an effect on the chemo-thermal evolution of collapsing clouds, depending on
the density $\rho$, gas temperature $T$, density $\rho _i$ of a grain species $i$, and 
$\rho _{X}$ of a key element $X$.
To reduce the computational cost, we have calculated the following quantities:
\begin{itemize}
\item[(1)] H$_2$ formation rate on grain surfaces from Eq. (4) of \citetalias{Chiaki15}
\item[(2)] dust opacity from Eq. (8) of \citetalias{Chiaki15}
\item[(3)] dust cooling rate from Eq. (7) of \citetalias{Chiaki15}
\item[(4)] grain growth rate (see below)
\end{itemize}
for each species $i$, and tabulated them.

As a result of grain growth, these rates are enhanced.
If the growth rate depends on the grain radius, it would be almost impossible to tabulate
these rates for each grain species because the size distribution is deformed over the course
of time.
We pay attention to the convenient nature of grain growth in that
the size increment is independent of the grain radius.
Although the rate at which a gas-phase species sticks to grain surfaces is proportional to $r ^2$,
it is converted to the increasing rate of grain radius with divided by a grain surface area ($\propto r^2$)
for the impactor to cover the entire area of a grain surface.
Because the size distribution function is just shifted to larger radii without changing its shape,
we can estimate the rates from the initial size distribution and
dust amount at a given time.

In \citetalias{Chiaki15}, the rates (1)--(4) are formulated with the size
increment $\delta r_i(t)$ for a grain species $i$ at the time $t$.\footnote{The origin of time
should be the time at which grain size distribution is fixed.
Here we set it the time of SN explosion.
Note that it takes $\sim 10^4$ yr that dust size distribution is fixed \citep{Nozawa07}.}
When we create the lookup table, we convert the given fluid quantities into $\delta r_i(t)$ in the following manner.
In the relevant density and temperature range,
metals are accreted onto pre-existing seed grains, i.e., seeds are not newly formed.
Therefore, the ratio of number density of grain particles\footnote{In Eq. (\ref{eq:ndust}),
$\varsigma _i$ is
the bulk density of grain species $i$.
$\langle \cdots \rangle _{i,t}$ 
is the average of a physical quantity weighted by
the size distribution function $\varphi _{i,t} (r)$ normalized as 
$\int \varphi _{i,t} (r) dr = 1$.}
\begin{equation}
n_i(t) = \frac{\rho _i(t)}{(4\pi/3) \langle r^3 \rangle _{i,t} \varsigma _i}
\label{eq:ndust}
\end{equation}
to that of nuclei of the key element is constant:
\begin{equation}
\frac{\rho _i(0) / \langle r^3 \rangle _{i,0}}{\rho _X (0)}
=
\frac{\rho _i(t) / \langle r^3 \rangle _{i,t}}{\rho _X (t)}.
\label{eq:conserv}
\end{equation}
Since $\delta r_i (t)$ is independent on grain radii, 
$\langle r^3 \rangle _{i,t}$ is expanded as
\begin{eqnarray}
\langle r^3 \rangle _{i,t} 
&=& \langle (r+\delta r_i(t))^3 \rangle _{i,0} \nonumber \\
&=& \langle r^3 \rangle _{i,0}
+ 3\langle r^2 \rangle _{i,0} \delta r_i(t) 
+ 3\langle r \rangle _{i,0} (\delta r_i(t))^2
+ (\delta r_i(t))^3, \nonumber \\
\label{eq:size_increment}
\end{eqnarray}
using the moments of the grain radii $\langle r^n \rangle _{i,0}$
($n=1,2,3$) weighted by the initial size distribution function $\varphi _{i,0} (r)$.
Then, $\delta r_i (t)$ is derived as the root of the cubic equation 
(\ref{eq:size_increment}).

We should note that $\delta r_i (t)$ is independent of $r$
unless grain sublimation is considered.
In our chemistry model, we ignore it because grains sublimate only after the gas becomes opaque
in the continuum \citep{Omukai00}.
Also, grain collisional cross-sections with gas-phase species
depend on $r$ when grains interact with ions and
electrons \citep{Draine87}.

Next, we describe how to estimate grain growth rate.
It is treated as a set of chemical reactions.
The grain growth rate is written as
\begin{equation}
k_{{\rm gg}, i}(t) = \min _x \left\{4\pi \langle r^2 \rangle _{i,t} v_{{\rm th}, x} n_i n_x\right\},
\end{equation}
where $v_{{\rm th},x}$ and $n_x$ are the thermal velocity and number density of a reactant $x$
bearing the corresponding element $X$, respectively.
For species with multiple reactants (e.g., Mg, SiO, and H$_2$O for forsterite),
we take the minimum rate among them \citep{Kozasa87}.
Then, the growth rate of number density $\nu _i$ of nuclei of the key element $X$ locked up into grain
species $i$ and 
the depletion rate of number density $n_x$ of gas-phase species $x$
are
\begin{equation}
 \left. \frac{{\rm d}\nu _i(t)}{{\rm d}t} \right| _{\rm gg} =
-\left. \frac{{\rm d}n_x(t)}{{\rm d}t} \right| _{\rm gg} 
= k_{{\rm gg}, i}(t) \prod _{x} n_x.
\end{equation}
The number density $\nu _i$ is converted into the mass density of grain species $i$ as
\begin{equation}
\rho _i (t) = \nu _i (t) \mu _i \mH,
\end{equation}
where $\mu _i$ is the molecular weight of a monomer of $i$.

\section{Comparison of chemistry with one-zone calculation}

\begin{figure*}
\includegraphics[width=5.7cm,natwidth=4inch,natheight=3.2inch]{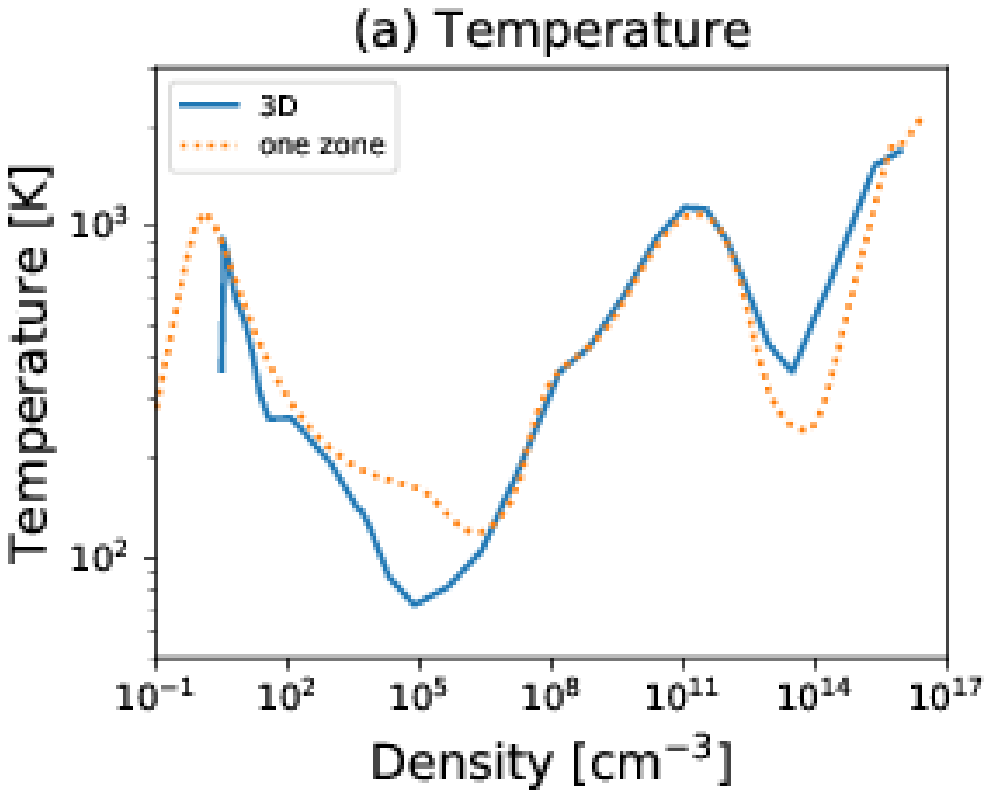}
\includegraphics[width=5.7cm,natwidth=4inch,natheight=3.2inch]{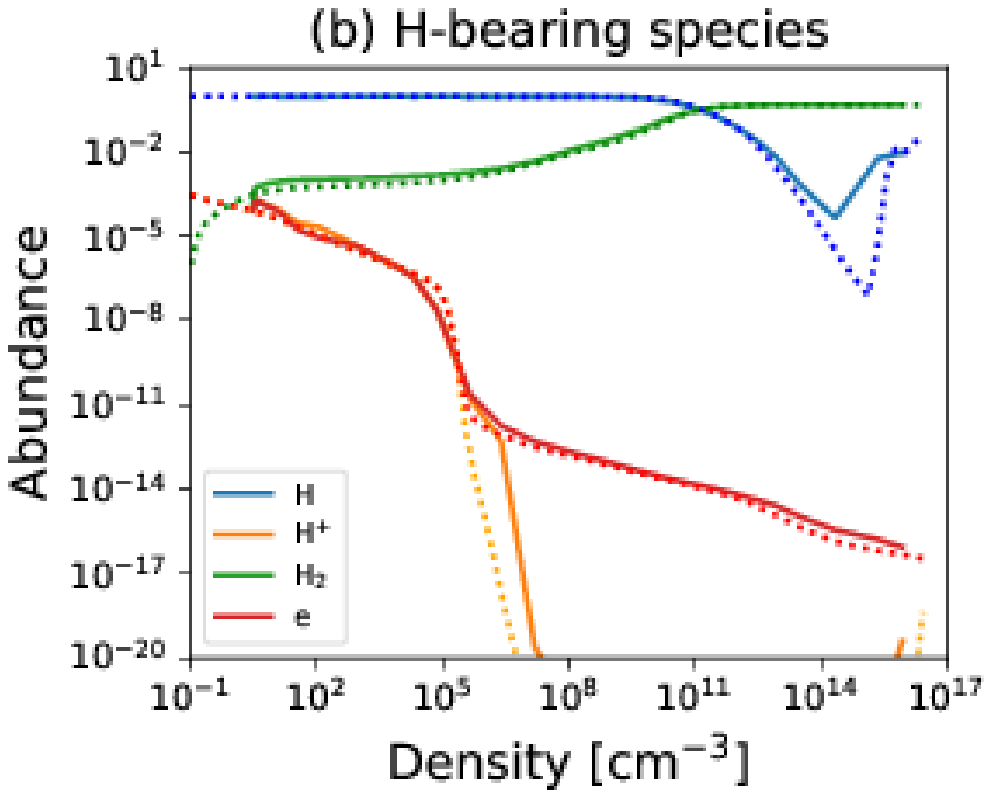}
\includegraphics[width=5.7cm,natwidth=4inch,natheight=3.2inch]{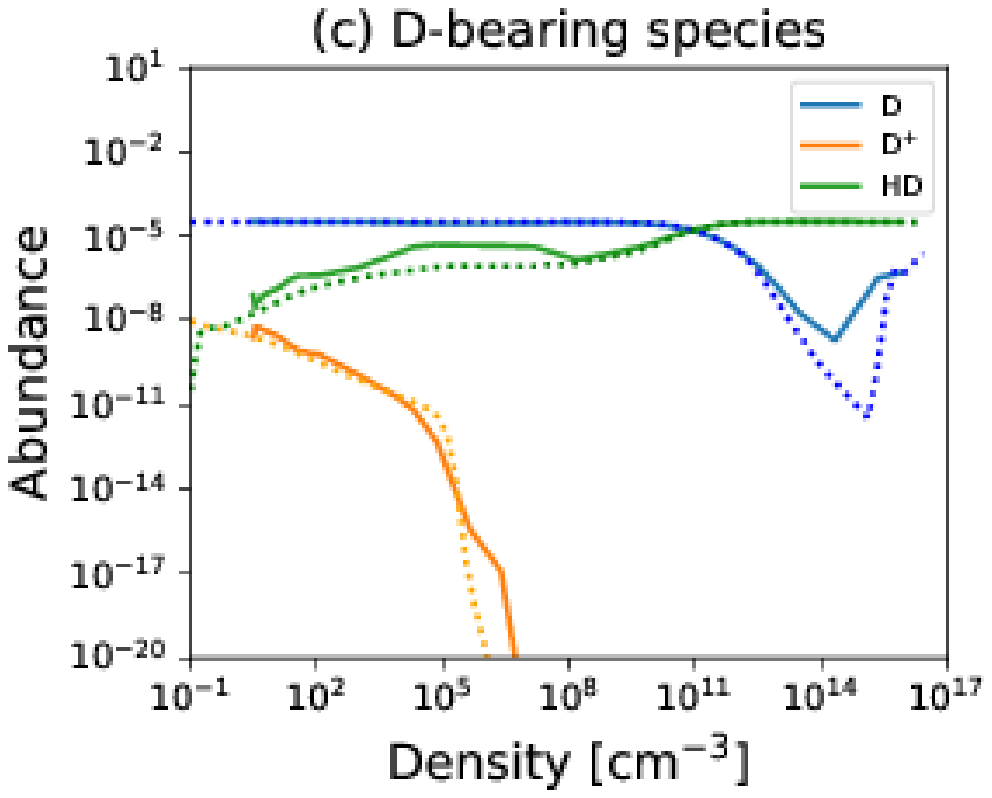}
\includegraphics[width=5.7cm,natwidth=4inch,natheight=3.2inch]{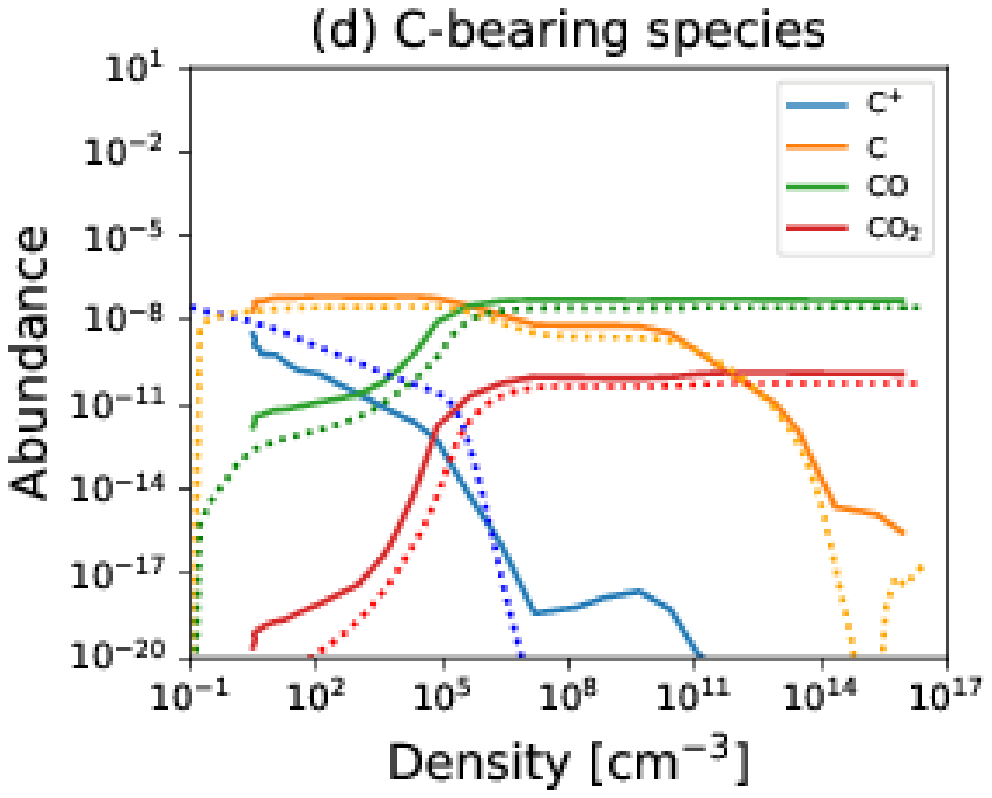}
\includegraphics[width=5.7cm,natwidth=4inch,natheight=3.2inch]{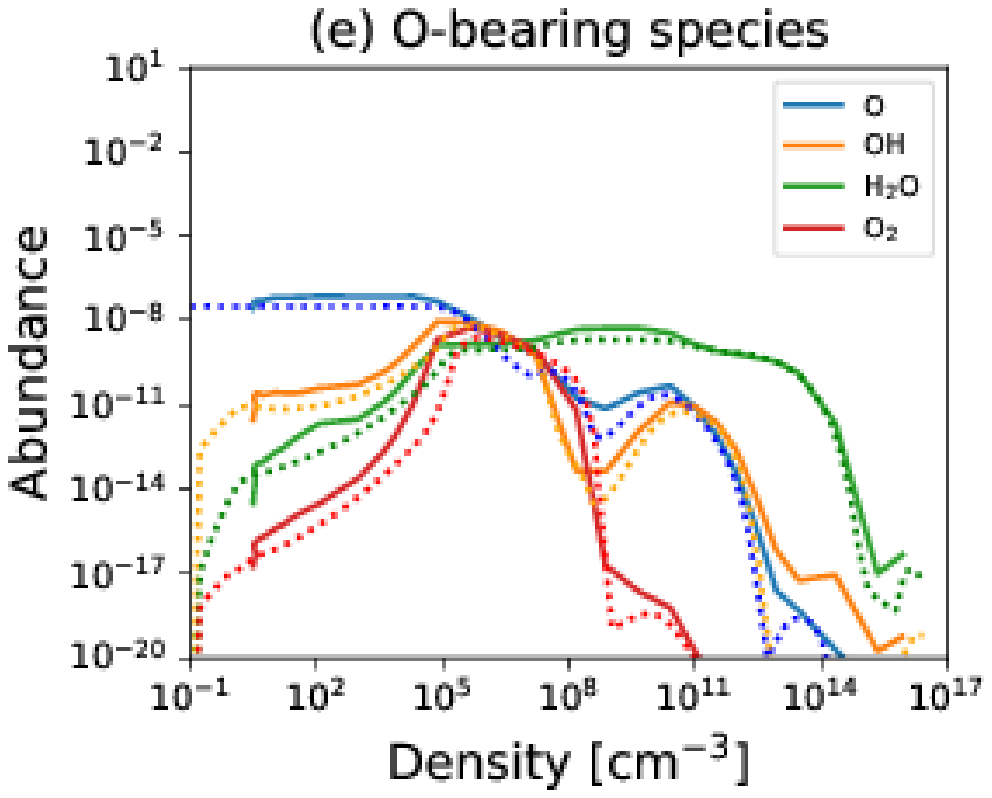}
\includegraphics[width=5.7cm,natwidth=4inch,natheight=3.2inch]{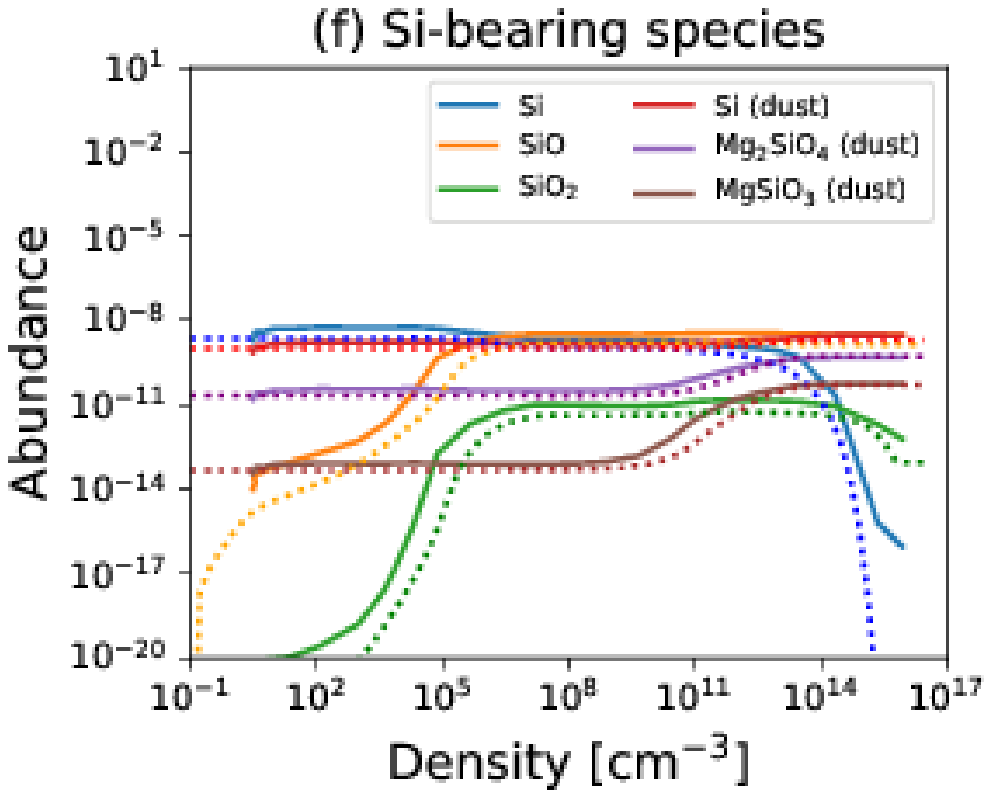}
\caption{
Same as Fig. \ref{fig:chemothermal_evolution} but for the major species in our 3D simulation (solid curves)
compared with the result of one-zone calculations (dotted curves).
}
\label{fig:chemothermal_evolution_comparison}
\end{figure*}

We compare our simulation results with a semi-analytic one-zone collapse calculation,
in which the thermal evolution of a collapse center is followed with the same chemical
reaction network and radiative cooling as the simulation \citep[see also][]{Omukai00}.
The density $\rho$ is assumed to increase at the rate of nearly a free-fall time $\tff = (3\pi/32G\rho)^{1/2}$:
\begin{equation}
\frac{{\rm d}\rho}{{\rm d}t} = \frac{\rho}{\tcol}.
\end{equation}
In this equation, $\tcol$ is the collapse timescale as
\begin{equation}
\tcol = \frac{\tff}{\sqrt{1-f_p}},
\label{eq:tcol}
\end{equation}
where $f_p$ is the ratio of gas thermal energy to gravitational energy as
\begin{eqnarray}
f_p = 
  \begin{cases}
    0, ~~~~~~~~~~~~~~~~~~~~~~~~~~~~~~~~~~~~~~~~~~~~~~(\gamma < 0.83) \\
    0.6 + 2.5(\gamma -1) -6.0(\gamma -1)^2, ~~~~~~~~~(0.83 < \gamma < 1) \\
    1.0 + 0.2(\gamma -4/3) - 2.9(\gamma -4/3)^2, ~~~(\gamma > 1) 
  \end{cases}
\end{eqnarray}
$\gamma = {\rm d}\ln p / {\rm d}\ln \rho$ is the specific heat ratio of the gas
with density $\rho$ and pressure $p$ \citep{Larson69, Omukai05}.
Then, specific energy $e$ is calculated as
\begin{equation}
\frac{{\rm d}e}{{\rm d}t} = -p\frac{{\rm d}(1/\rho)}{{\rm d}t} + \Gamma - \Lambda,
\end{equation}
where $\Gamma$ and $\Lambda$ is the gain and lost of the energy, respectively.

Fig. \ref{fig:chemothermal_evolution_comparison} shows the evolution of temperature and 
chemical abundances with a metallicity $Z = 2.6\E{-4} \ \Zsun$, the same as the recollapsing
region in our simulation.
We find that our three-dimensional results match very well with the one-zone model.
However, there is the discrepancy at densities $\nH < 10^7 \ \percc$ and $\nH > 10^{11} \ \percc$.
This difference occurs because the one-zone calculation does not include any hydrodynamic effects.
In our simulation, the collapse timescale $\tcol$ is longer than the one 
in Eq. (\ref{eq:tcol}).
At $\nH < 10^7 \ \percc$, the collapse time is slower than in Eq. (\ref{eq:tcol})
because of the gravitational potential of the dark matter MH whose distribution is more
spread than gas \citepalias[see also][]{Chiaki16}.
Since the adiabatic compressional heating rate $\Gamma _{\rm ad} = p/\tcol$ is smaller and
thus the chemical timescale for H$_2$ formation is relatively longer, 
the gas temperature is lower than the one-zone calculation.
Also, when temperature decreases below 150 K, HD molecule formation is enhanced, and
the gas further cools down to 70 K. 
Conversely, at $\nH > 10^{11} \ \percc$, the temperature in our simulation is higher than
one in the one-zone calculation.
The collapse timescale is shorter than the one in the one-zone calculation (Eq. (\ref{eq:tcol})) because of
the rapid gas accretion into the hydrostatic core formed by H$_2$ formation heating 
(see Sec. \ref{sec:thermal_evolution}).


\label{lastpage}

\end{document}